\documentclass[10pt,journal,compsoc]{IEEEtran}

\usepackage{amsmath}
\usepackage{amssymb}

\usepackage{bbold}
\usepackage[export]{adjustbox}
\usepackage{authblk}

\usepackage{hhline}
\usepackage{array}
\ifCLASSOPTIONcompsoc
  \usepackage[nocompress]{cite}
\else
  \usepackage{cite}
\fi

\usepackage{url}
\usepackage{balance}
\usepackage{bm}

\usepackage{graphicx}

\usepackage{xcolor}
\usepackage{amssymb}
\usepackage{pifont}

\newcommand{\mourmeth}{\texttt{LinksIQ}}
\newcommand{\ourmeth}{$\mourmeth$\xspace}
\newcommand{\miq}{\texttt{IQ}}
\newcommand{\iq}{$\miq$\xspace}

\newcommand{\eat}[1]{}

\newcommand{\vsa}{\vspace*{-0.28cm}}

\newcommand{\vsc}{\vspace*{-0.4cm}}

\usepackage{cleveref}

\usepackage{enumitem}
\setlist[itemize]{leftmargin=*}

\usepackage[linesnumbered,ruled]{algorithm2e}
\usepackage{ifpdf}
\usepackage{eqparbox}

\usepackage{subcaption}

\usepackage{fixltx2e}
\usepackage{stfloats}

\usepackage{multirow}

\begin{document}
%
\title{\textbf{\ourmeth}: Robust and Efficient Modulation Recognition with Imperfect Spectrum Scans}

\author[]{Wei Xiong}
\author[]{Karyn Doke}
\author[]{Petko Bogdanov}
\author[]{Mariya Zheleva}
\affil[]{Department of Computer Science, University at Albany SUNY, \\ \textit{\{wxiong, kdoke, pbogdanov, mzheleva\}@albany.edu}}

\IEEEtitleabstractindextext{%

\begin{abstract}
Modulation recognition plays a key role in emerging spectrum applications including spectrum enforcement, resource allocation, privacy and security. 
While critical for the practical progress of spectrum sharing, modulation recognition has so far been investigated under unrealistic assumptions: (i) a transmitter's bandwidth must be scanned alone and in full, (ii) prior knowledge of the technology must be available and (iii) a transmitter must be trustworthy. In reality these assumptions cannot be readily met, as 
a transmitter's bandwidth may only be scanned intermittently, partially, or alongside other transmitters, and modulation obfuscation may be introduced by short-lived scans or malicious activity. 

This paper bridges the gap between real-world spectrum sensing and the growing body of methods for modulation recognition designed under simplifying assumptions. To this end, we propose to use local features, besides global statistics, extracted from raw \iq data, which collectively enable \ourmeth, a robust framework for modulation recognition with imperfect spectrum scans. Our key insight is that ordered \iq samples from spectrum traces form distinctive patterns across modulations, which persist in the face of spectrum scan deficiencies. 
We mine these patterns through a Fisher Kernel framework that captures the non-linearity in the underlying data.  
With these domain-informed features, we employ lightweight linear support vector machine classification for modulation detection. 
Our framework is robust to noise, partial transmitter scans and data biases 
without utilizing prior knowledge of the underlying transmitter technology. The recognition accuracy of our approach consistently outperforms baselines in both simulated and real-world traces. 
We evaluate and compare the performance of our approach in a controlled testbed using two popular software-defined radio platforms, RTL-SDR and USRP. We demonstrate high detection accuracy (i.e. 0.74) even with a \$20 RTL-SDR scanning at 50\% transmitter overlap. This constitutes an average of 43\% improvement over existing counterparts employed on RTL-SDR scans. We also explore the effects of platform-aware classifier training and discuss implications on real-world modrec system design. Our results demonstrate the feasibility of low-cost transmitter fingerprinting at scale. 
\end{abstract}

\begin{IEEEkeywords}
Modulation classification, cognitive radio, machine learning, feature engineering, local patterns.
\end{IEEEkeywords}

} 

\maketitle

\IEEEdisplaynontitleabstractindextext

%
\IEEEpeerreviewmaketitle

\section{Introduction}\label{sec:intro}

In the past decade, underpinned by the rapid growth of wireless communications demand, a plethora of emerging communication technologies have been employed from TV White Spaces~\cite{tvws} and Citizens Broadband Radio Service (CBRS)~\cite{cbrs} to visible light communications. While these developments bring hope for fast and high-quality last mile connectivity, they also demand superior spectrum resources. 
In response, Dynamic Spectrum Access (DSA) has emerged as a promising solution, which allows the opportunistic allocation of spectrum resources on demand. To adopt DSA, wireless devices are required to have constant cognizance of the spectrum availability and quality and employ these insights in agile utilization of the underlying radio resources. 
The success of such opportunistic access, however, critically hinges on devices' ability for autonomous spectrum characterization and transmitter fingerprinting.

Modulation recognition (modrec) is a key transmitter fingerprinting task of critical importance to both civil and defence applications~\cite{nandi1997modulation}. Modulation recognition in practice consists of a two-step process: data collection (i.e. spectrum sensing) and data analysis (i.e. recognition). While the quality and quantity of collected data inevitably affects the recognition accuracy, existing modrec approaches are largely disconnected from the underlying spectrum sensing techniques that generate the data necessary for analysis.
This disconnect will further widen with the advent of autonomous spectrum sensing and agile transmitter technology. \textbf{\textit{Future spectrum sensing infrastructures}} will have to leverage multiple dedicated~\cite{so_ms,roy2017cityscape,McHenry} or crowdsourced~\cite{chakraborty2017specsense,chakraborty2018spectrum,Nika:2016:EVC:2994551.2994557} spectrum sensors, collecting traces in a wide frequency band. To support this heterogeneous environment, the sensor infrastructures will have to sequentially ``step through'' the spectrum, while collecting data from contiguous sub-bands~\cite{Nika:2016:EVC:2994551.2994557,Chakraborty:2016:BRU:2980115.2980129}. As a result, individual transmitter's activity will be scanned intermittently, with partial coverage of their occupied frequency band and alongside other transmitters or unoccupied spectrum sub-bands. 
\textbf{\textit{Modrec data analysis}} is in essence a classification problem approached via various machine learning techniques from lightweight support vector machines (SVM)~\cite{gang2004study} to artificial neural networks~\cite{nandi1997modulation}. Of key importance to the detection speed and accuracy are the features employed for classification which are extracted from raw measured \iq samples. The state-of-the-art feature-based methods employ two families of features: \emph{order statistics (OS)}~\cite{han2017low} and \emph{high order cumulants (HOC)}~\cite{swami2000hierarchical,dobre2003higher,lu2017modulation}. The former family employs sorted \iq sample components for classification, while the latter extracts high-order statistics from the distribution of samples. We refer to HOC and OS as \textit{global} features, since they utilize global statistics over multiple samples. Thus, \emph{both families of global features disregard the sequential order of \iq samples, effectively treating them as a ``bag'' of independent values. We demonstrate that the information encoded in this local sequential order is reflective of the underlying modulation and show how it can be leveraged for robust and practical modulation recognition.}

All prior work in modulation recognition poses prohibitively-high spectrum sensing requirements closely resembling the steps necessary for signal decoding. First, a transmitter of interest has to be scanned alone and in full, whereby the sensor's center frequency and bandwidth have to be aligned with these of the transmitter and all side-band signals have to be filtered out~\cite{han2017low, swami2000hierarchical, dobre2003higher, lu2017modulation}. 
In addition, transmitters should be scanned for a sufficiently long duration, such that each modulation symbol is uniformly represented in the collected trace. 
However, emerging spectrum sensing systems increasingly challenge these modrec requirements, as they perform sweep-based wideband spectrum scans, and introduce intermittency, partiality and biases in a given transmitter's scan. 
In our preliminary analysis, we observe that relaxing the stringent sensing requirements imposed by existing HOC- or OS-based modrec approaches leads to severe deterioration in the classification performance. In \S\ref{sec:limitations}, we demonstrate a significant sensitivity of HOC and OS features to scan partiality, data bias, and constellation rotation. In particular, individual HOC and OS features converge across modulation types, while their standard deviations increase. These trends cause a dramatic reduction of their discriminative power, which, in turn, leads to poor modrec performance regardless of the utilized classifier. \emph{The gap between future spectrum sensing requirements and the assumptions of existing modrec methodology calls for novel data-driven approaches for robust modulation recognition in the face of partial, intermittent, biased or noisy scans.}

To bridge this gap, we design \ourmeth, a framework that leverages novel features from local patterns in \iq samples for robust modulation recognition with partial, biased and noisy scans. Specifically, we use the phase and amplitude of \iq samples to create ordered subsequences of values, dubbed \textit{shingles}. We represent an \iq sample sequence in terms of its shingles based on a Fisher Kernel generative framework~\cite{perronnin2010improving}, where we quantify gradient statistics for shingles being generated by a Gaussian Mixture Model (GMM) dictionary of prototypical shingles. We train and employ SVM~\cite{cortes1995support} classifier for run-time detection of a transmitter's modulation without prior knowledge of the scan's partiality, transmitter technology, data bias or the channel signal to noise ratio (SNR). We note that SVM is simply one classifier choice; while our primary focus is on robust feature design, the \ourmeth framework is extensible to other classifiers including from the artificial neural networks family. We demonstrate robust performance of our method in both a realistic MATLAB simulation and a testbed  comprised of controlled USRP-based transmitters and heterogeneous USRP- and RTL-based sensors. 

Our paper makes the following key contributions: \\
\noindent $\bullet$ We are the first to conceptualize the problem of modulation recognition from partial spectrum scans.\\
\noindent $\bullet$ We are the first to propose and demonstrate the potential of local \iq patterns for modulation recognition in future spectrum sensing platforms. \\
\noindent $\bullet$ We design \ourmeth, a modrec framework that includes an adaptive Fisher Kernel feature extractor and a lightweight SVM classifier for robust modrec from \iq sequence patterns. \\
\noindent $\bullet$ \ourmeth exhibits a significant improvement of modrec accuracy over baselines in both realistic simulation and real-world testbed spectrum measurement.\\
\noindent $\bullet$ We demonstrate the feasibility of modrec with \$20 RTL-SDRs performing partial scans and outline prospects for low-cost classifier training towards ubiquitous modrec. 

\vspace{-.3cm}

\section{Related work} \label{sec:related}
Our related work falls in two categories. The first one pertains to existing modulation recognition literature, while the second draws from other domains that use local sequential patterns as classification features. We discuss these in turn.

\noindent \textbf{Modulation recognition} has been an active area of research with two main streams of methodology: likelihood-based (LB)~\cite{panagiotou2000likelihood} and feature-based (FB)~\cite{dobre2007survey}. While optimal, LB approaches suffer high computational complexity and are not resilient to RF chain imperfections (e.g. timing and frequency offset), and wireless channel effects (e.g. non-Gaussian noise)~\cite{swami2000hierarchical}. In addition, LB approaches explicitly rely on a model modulation constellation, which is not always readily available or may be significantly distorted due to small scan overlap with the transmitter, missing or unbalanced constellation symbols and high noise regimes. FB approaches offer a lower complexity alternative and have been heavily utilized in recent modrec literature~\cite{han2017low,swami2000hierarchical,dobre2003higher,lu2017modulation,abuella2016automatic,gang2004study}. FB modrec extracts features from measured \iq data and performs modulation classification based on these features. The state-of-the-art techniques adopt order statistics (OS)~\cite{han2017low}, high order cumulants (HOC)~\cite{swami2000hierarchical, dobre2003higher, gang2004study} and kernel density functions~\cite{abuella2016automatic} as features and employ various classification techniques including support vector machines~\cite{gang2004study} and artificial neural networks~\cite{nandi1997modulation}.  
All the above approaches pose unrealistic requirements to spectrum sensing including $100\%$ transmitter scan overlap with the transmitter, side band exclusion, and are sensitive to the signal's noise level. All methods except~\cite{lu2017modulation} assume no bias in symbol representation. These requirements are in direct disagreement with future spectrum sensing infrastructures, which will use dedicated~\cite{so_ms,roy2017cityscape,McHenry} or crowdsourced~\cite{chakraborty2017specsense,chakraborty2018spectrum,Nika:2016:EVC:2994551.2994557} sensors for wideband intermittent sensing in support of spectrum sharing technology, policy and enforcement. Lu et al.~\cite{lu2017modulation} consider modulation recognition from incomplete and biased scans, however, the method employs HOC features which, as we demonstrate in \S\ref{sec:background} (i) are highly-sensitive to scan imperfections, (ii) but encode complementary information to our proposed local features, and thus can be successfully combined in order to boost modrec performance (see \S\ref{sec:evaluation}). 
Recently, deep neural networks (DNN) have been employed for modrec with promising performance outcomes~\cite{o2018over,rajendran2018deep}. Such approaches are orthogonal to our work, as they use simple input data comprised of raw \iq samples while employing complex classifiers. In contrast, we focus on domain-informed feature design, and in this paper, employ lightweight SVM classification, however, our features can be employed in a DNN framework as well. 

\noindent \textbf{The discriminative power of local patterns} has been demonstrated in various signal domains, including images~\cite{zhang2010discriminative,gao2010local,sanchez2013image,zhang2007local,yu2010improved}, video~\cite{mironica2013fisher}, audio~\cite{kumar2016weakly} and text~\cite{cavnar1994n}. The-state-of-the-art techniques employ dictionary learning for feature extraction and various classifiers for classification~\cite{zhang2010discriminative,sanchez2013image}. A key benefit of local patterns is their resilience to global changes in the underlying signal. As demonstrated in \S\ref{sec:background}, these benefits carry over in the modulation recognition domain, where the misrepresentation of a constellation symbol, higher noise level or partial transmitter overlap inevitably affect global features, but preserve inherent signatures in local \iq patterns. 

\vspace{-.3cm}

\section{Preliminaries}
\label{sec:background}

In this section we discuss some preliminaries that underpin our work. We begin by describing the gap between modrec requirements and sensing capabilities. We then present the modrec signal model and corresponding HOC and OS features from the literature. Finally, we empirically evaluate the limitations of existing features with scan imperfections.

\vspace{-.3cm}
\subsection{Sensing requirements for modulation recognition}

\begin{figure}[t]
\begin{minipage}{\linewidth}
\centering
\begin{subfigure}[b]{\textwidth}
\includegraphics[width=\textwidth]{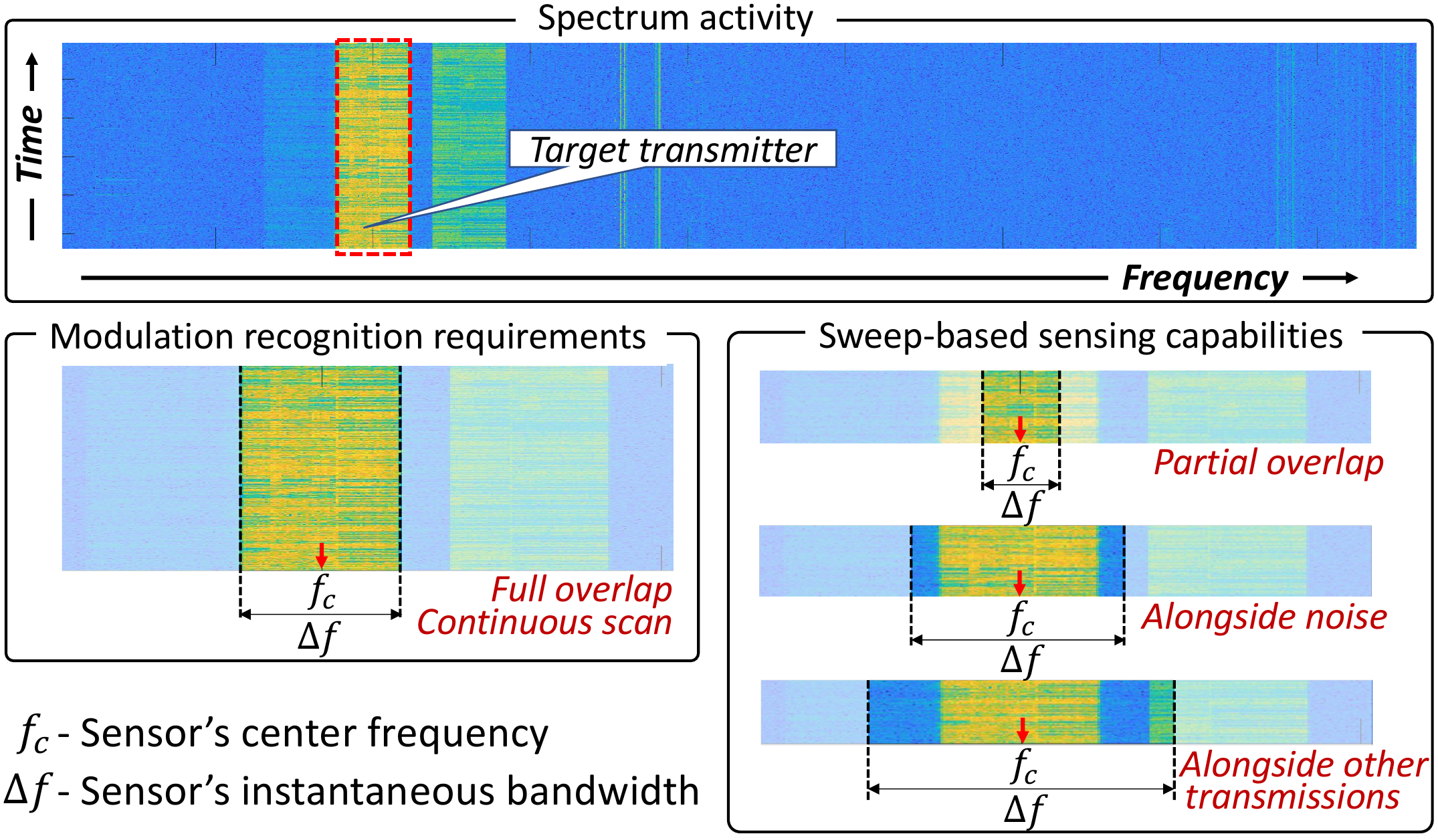}
\end{subfigure}
\vsa \vsa
\caption{\footnotesize The gap between modulation recognition requirements and emerging spectrum sensing capabilities requires novel modulation recognition approaches.}\label{fig:sensing_app}
\end{minipage}
\end{figure}

Existing modulation recognition methods have stringent sensing requirements, which are exemplified in Fig.~\ref{fig:sensing_app}. Imagine a transmitter of interest, as illustrated in the wideband scan on the top of the figure. Existing modrec techniques require that the transmitter's bandwidth is scanned at a 100\% overlap, excluding any side-band noise or transmissions (e.g. Fig.~\ref{fig:sensing_app} bottom left). In addition, the transmitter should be observed for a sufficiently long time, such that the resulting scan contains a uniform representation of the constellation symbols (see Fig.~\ref{fig:uniform} for an example of a uniform constellation representation). Existing methods assume that the transmitter is trustworthy in that it would not tamper with the transmitted data to introduce constellation biases. Finally, existing methods assume prior knowledge of a constellation's rotation and the channel properties. 

The above requirements may not be readily met by emerging spectrum measurement systems. 
The target band of spectrum sensing is typically in the order of several GHz~\cite{so_ms, McHenry, roy2017cityscape, van2017electrosense}, which can be two to three orders of magnitude larger than the instantaneous bandwidth of a spectrum sensor~\cite{rtlsdr,usrp}\footnote{For example, the maximum stable instantaneous bandwidth of a RTL-SDR~\cite{rtlsdr} is 2MHz, whereas that of a USRP~\cite{usrp} is 20-30MHz, depending on the model.}. As a result, spectrum measurement systems perform sweep-based sensing, stepping through the target bandwidth according to their set instantaneous bandwidth and dwelling on each subband for a pre-determined time interval dubbed \textit{dwell time}. This operation, as illustrated in the bottom right corner of Fig.~\ref{fig:sensing_app}, is poised to introduce various imperfections in the spectrum scans that directly contradict the sensing requirements of existing modrec approaches. With the sweep-based approach, a transmitter may be scanned partially, alongside side-band noise or other transmitters. The scans might be intermittent, leading to biases in the symbol representation. Finally, no prior assumptions can be made about the underlying transmitter technology. 

Our work in this paper focuses on the first case of scan imperfections: intermittently scanned transmitters with partial overlap between the sensor's and the transmitter's bandwidth. As we will soon demonstrate (\S\ref{sec:limitations} and \S\ref{sec:evaluation}), these effects have detrimental impact on the discriminative power of modulation classification features and substantially deteriorate the classification performance.

\vspace{-.4cm}
\subsection{Signal model}
The input to modulation recognition is a set of \iq samples represented as complex numbers of the form $I~+~iQ$, collected by a sensor at a specified center frequency and bandwidth. We transform each sample into (amplitude, phase) pairs $x=(A,\phi)= (\sqrt{I^2+Q^2},\arctan\frac{Q}{I})$. Let $x=((A_1,\phi_1),(A_2,\phi_2)\dots,(A_n,\phi_n))$ denote an ordered sequence (series) of samples to which we will also refer as an instance. Given a set of such instances $X=[x^{(1)},x^{(2)},\dots x^{(m)}]$ and the corresponding modulation types employed by the underlying sampled transmitters $y=[y^{(1)},y^{(2)},\dots, y^{(m)}]$, the objective in supervised modulation recognition is to learn a classifier $f(x)=y$, which can predict the modulation type of newly observed instances. A majority of the existing feature-based techniques (including ours) do not work directly with the samples $x$ to learn a classifier, but instead extract features from them which are then employed for classification. 

\vspace{-.4cm}
\subsection{Global features: order statistics and cumulants}
All existing methods treat samples within an instance $x$ as independent, and extract features that summarize their statistical properties. There are two main classes of such features: order statistics and high order cumulants, both aiming to summarize the overall distribution of all instance samples. Thus, we refer to them as global features. 

\noindent \textbf{Higher order cumulants (HOCs)~\cite{aslam2012automatic}.} This approach seeks to summarize the statistical properties of the \iq samples using high order complex cumulants~\cite{eriksson2009statistics}. Within this framework the instance observations are modeled as samples from a complex-valued stationary random process $x(n)$ and high-order cumulants associated with the empirical distribution are estimated and used as predictive features~\cite{swami2000hierarchical}. Subsets of the fourth-order $\{C_{40},C_{41},C_{42}\}$ and sixth-order cumulants $\{C_{60},C_{61},C_{62},C_{63}\}$ have received most attention in the modrec literature~\cite{swami2000hierarchical,gang2004study,dobre2003higher,aslam2012automatic}. These quantities are defined in terms of estimates of moments associated with the empirical \iq sample observations. For example, $C_{42}$ is defined as:
\begin{equation}
C_{42}=M_{42}-|M_{20}|^2-2M^2_{21},
\label{eq:cumulant}
\end{equation}
where $M_{kv}=E \lbrack x(n)^{k-v}x^{*}(n)^{v} \rbrack$ are the empirical estimates of the moments associated with the stationary process from which the \iq samples are drawn, and $x^{*}(n)$ denotes the complex conjugation of $x(n)$. We omit the exhaustive definition of all the above cumulants due to space limitations and refer the reader to~\cite{aslam2012automatic} for details.   

To remove the effect of the signal scale on cumulants, they are typically normalized by $C_{21}$~\cite{swami2000hierarchical}:
$
\hat{C}_{kv} = C_{kv} / (C_{21})^{k/2}. 
$ 
In addition, since some cumulants are complex numbers, their $L_2$ is adopted as a real feature in classification~\cite{gang2004study}.

\noindent \textbf{Order statistics (OS)~\cite{han2017low}.} The $k$-th order statistic of a random real sample is its $k$-th smallest value. This simple notion gives rise to a modrec approach proposed in~\cite{han2017low} which employs the ordered values of the amplitude $A$, phase $\phi$ and the baseband \texttt{I} and \texttt{Q} components derived from an observed sample sequence $x$. OS features offer an alternative global summary of the distribution of the \iq samples. Note that in this representation the order of \iq samples is lost, however, as we demonstrate, this order contains information that can be used to discriminate modulations in realistic scenarios. 

\vspace{-.4cm}
\subsection{Limitations of global feature approaches}
\label{sec:limitations}
While the two families of global features discussed above have been successfully employed by many recent modrec approaches, they inherently rely on assumptions about (i) the overlap of the sensing range with the underlying transmitter's frequency range; (ii) the balance of observed symbols in a sample; and (iii) the phase offset (or constellation rotation). When these assumptions are relaxed in practical modrec ``in the wild'', the discriminative power of the global features deteriorates. In what follows, we study the robustness of HOC and OS to scan partiality, symbol biases and constellation rotation in order to quantify and understand their limitations.

\begin{figure}[t]
\begin{minipage}{\linewidth}
\centering
\includegraphics[width=.6\linewidth]{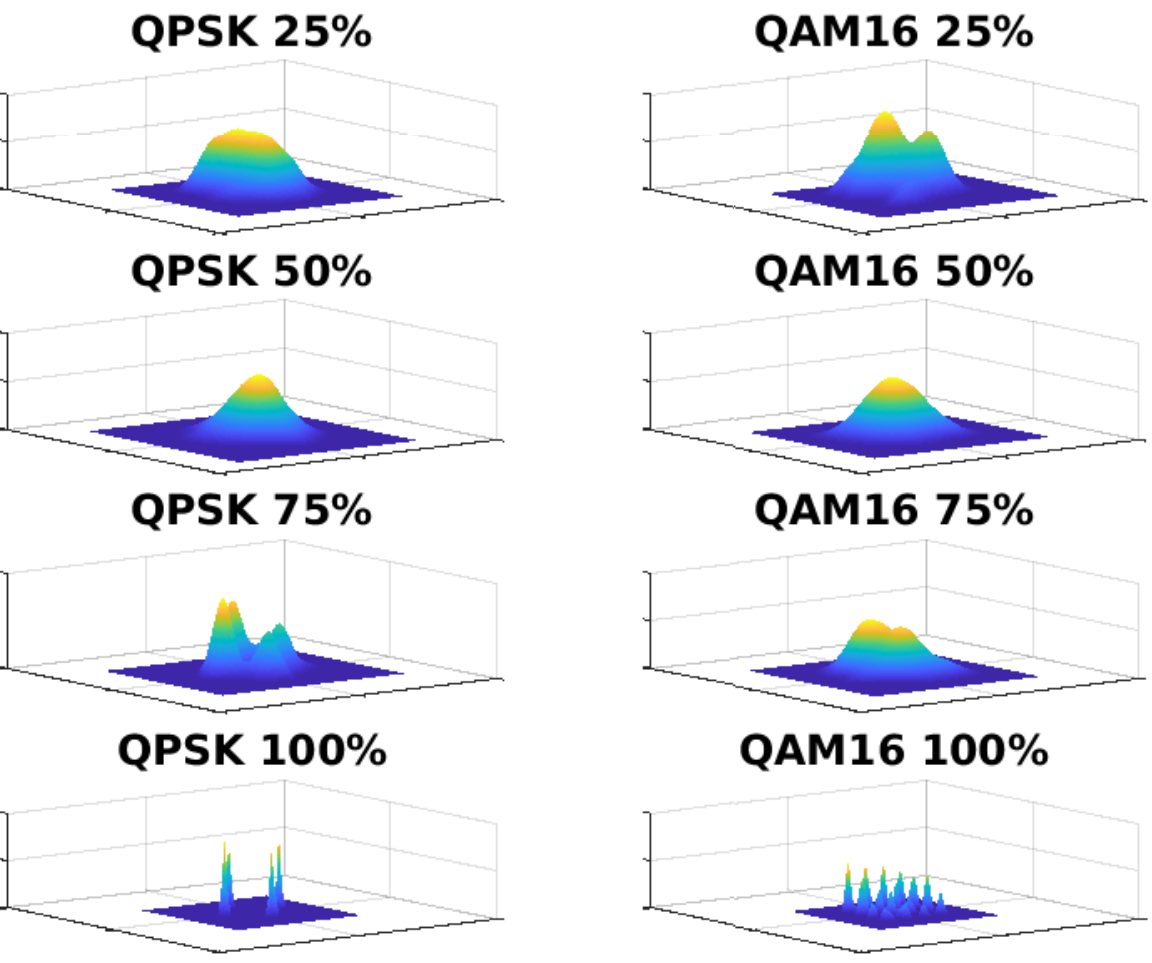}
\vsa
\caption{\footnotesize Effects of scan partiality on constellation shape for QPSK (left) and 16-QAM (right). 
}\label{fig:partBandEffects}
\vsa
\end{minipage}
\end{figure}

\noindent {\bf 1) Effects of partial scan overlap with the transmitter.} In sweep-based spectrum sensing, a transmitter may only be scanned partially as the exact frequency range may not be available a priori. 
Thus, we ask \emph{What is the effect of scan partiality on the shape of the modulation constellation, and in turn, on the discriminative power of global features?} Fig.~\ref{fig:partBandEffects} illustrates qualitatively this effect for \texttt{QPSK} and \texttt{16-QAM}. The constellations in both cases transition from sets of well-pronounced symbol clusters at $100\%$ overlap to fewer high-variance clusters at lower overlap, from which the original constellation is hard to recover. 

\begin{figure}[t]
\centering
\begin{subfigure}[t]{0.49\linewidth}
    \centering
    \includegraphics[width=\linewidth]{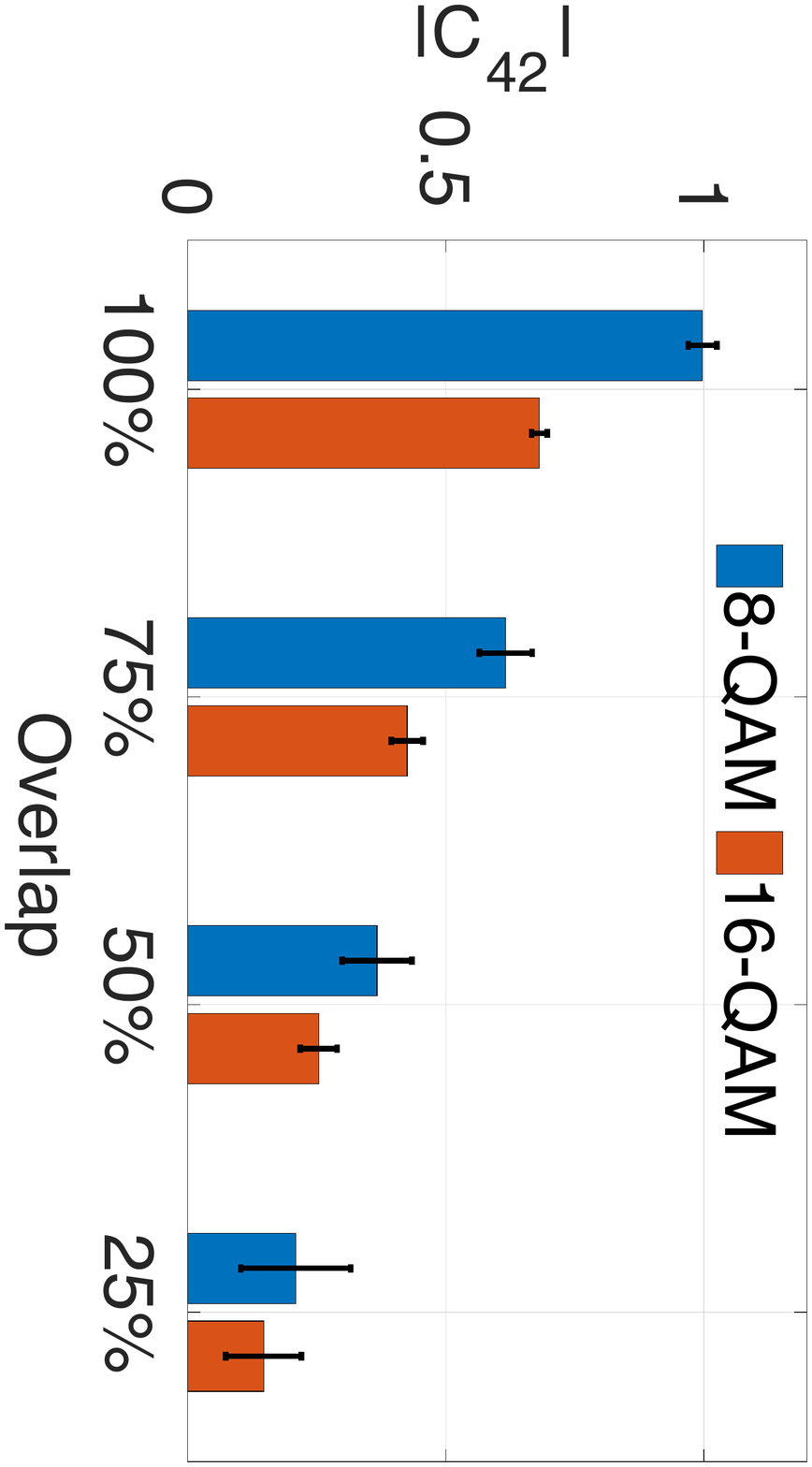}
    \caption{\footnotesize $C_{42}$ (partial overlap)}
    \label{fig:c42partial}
\end{subfigure}
\begin{subfigure}[t]{0.49\linewidth}
    \centering
    \includegraphics[width=\linewidth]{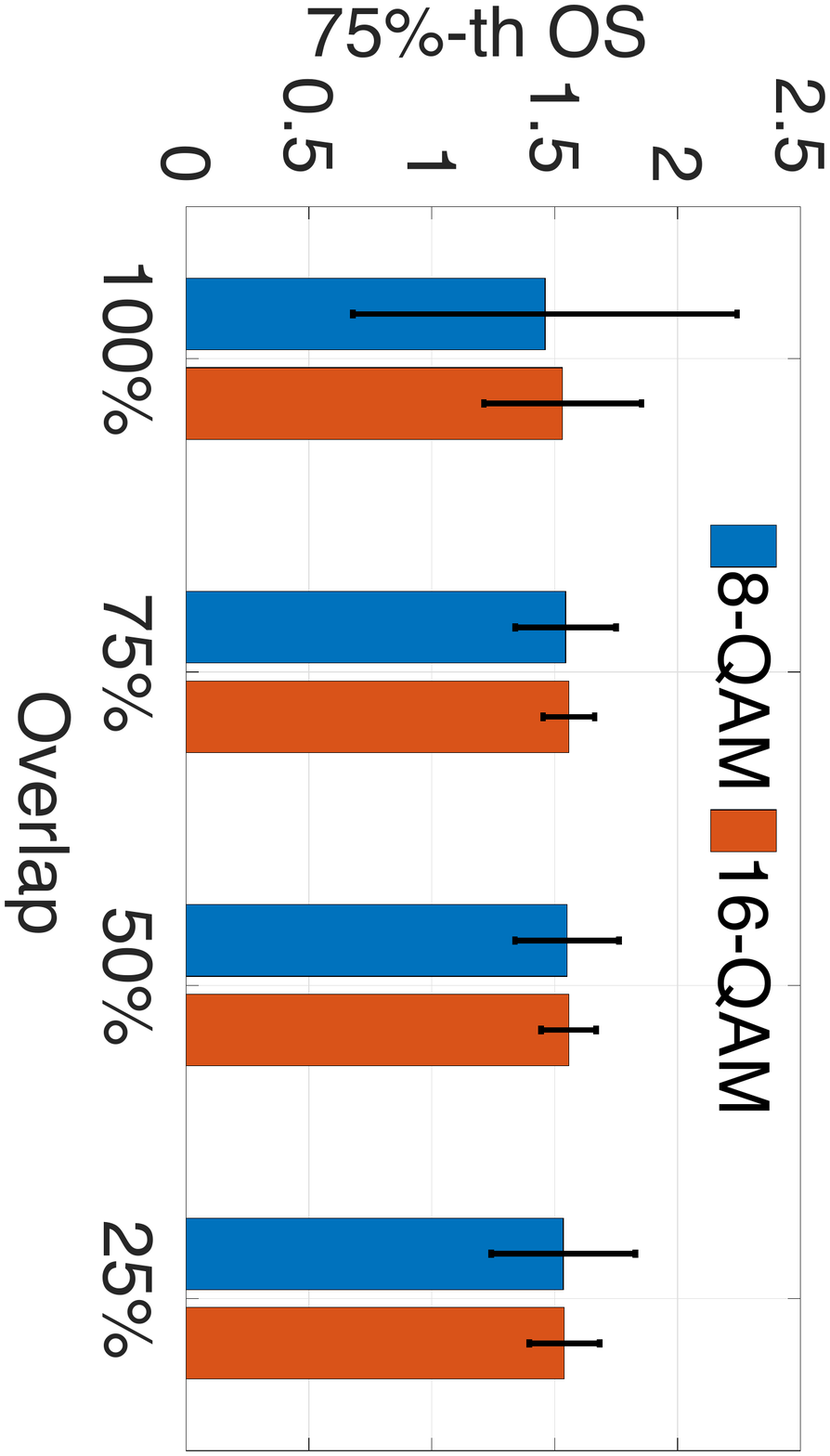}
    \caption{\footnotesize $75\%$ OS (partial overlap)}
    \label{fig:ospartial}
\end{subfigure}
\vsa
\caption{\footnotesize Effects of partial scans on global features' discriminative power. Y-axis presents the mean and standard deviations over $100$ instances of (\subref{fig:c42partial}) $C_{42}$ HOC and (\subref{fig:ospartial}) the $75\%-$ile OS with decreasing overlap.}
\label{fig:global-stability-partial}
\end{figure}

This visual deterioration of the constellation shape results in decreased stability of HOC and OS features across instances, which we study in Fig.~\ref{fig:global-stability-partial}. We generate 100 instance of 128 \iq samples each, and calculate the $C_{42}$ HOC and the $75\%$-ile OS. The figure presents the average and standard deviation of these statistics for two sample modulations (\texttt{8-QAM} and \texttt{16-QAM}) with decreasing overlap of the sensor's bandwidth with that of the transmitter. For $C_{42}$ (Fig.~\ref{fig:c42partial}), the averages across \texttt{8-QAM} and \texttt{16-QAM} converge, while their standard deviations increase. The same trend is observed for the $75\%$-th order statistic (Fig.~\ref{fig:ospartial}) and for other HOC and OS features (omitted in interest of space). As a result the discriminative power of global features decreases with lower overlap, which in turn, reduces its utility regardless of the adopted classifier.

\begin{figure}[t]
\begin{minipage}{\linewidth}
\centering
 \begin{subfigure}[t]{0.28\linewidth}
 \includegraphics[width=\textwidth,height=2cm]{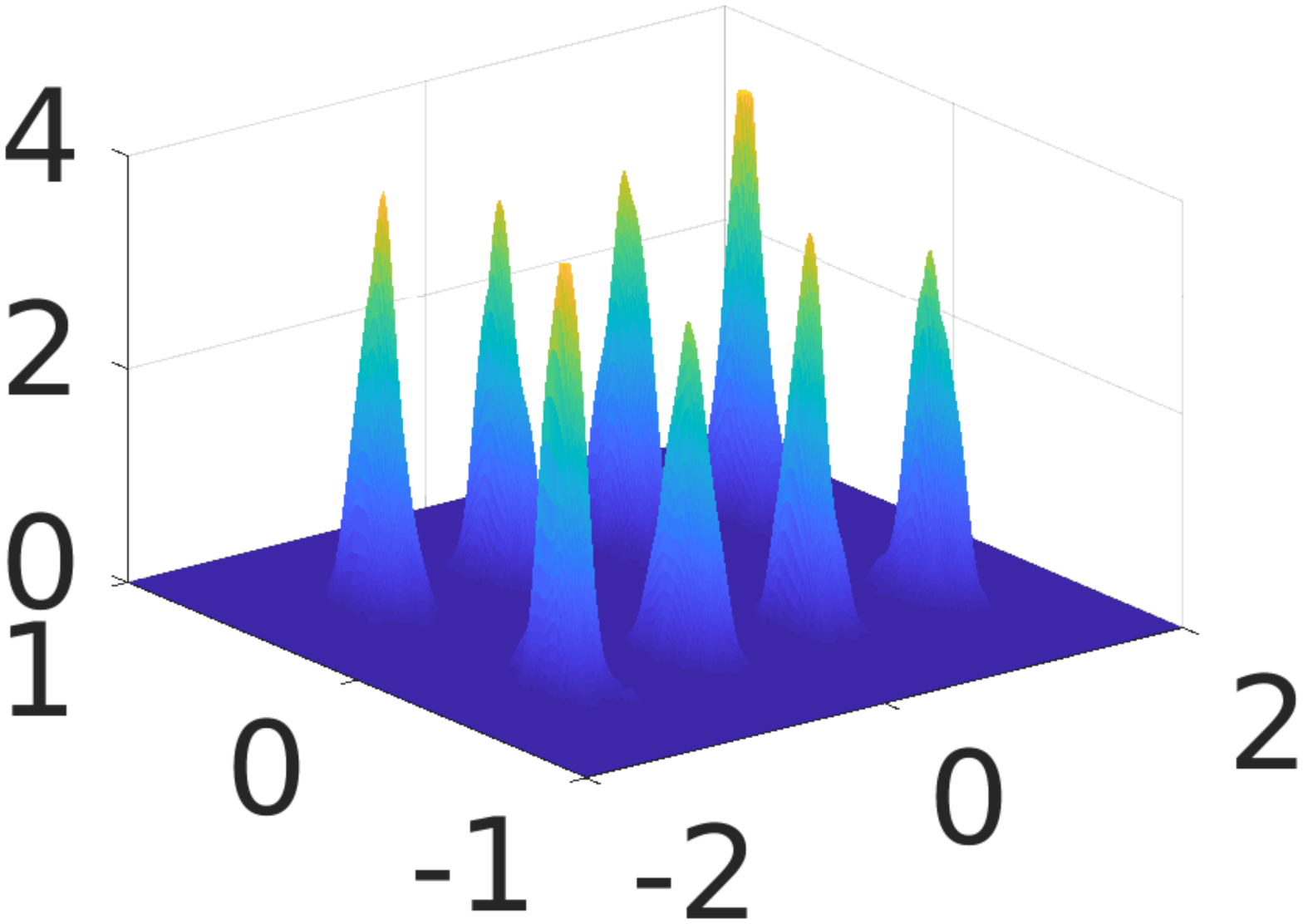}
 \caption{Uniform}\label{fig:uniform}
 \end{subfigure}
 \hfill
 \begin{subfigure}[t]{0.28\linewidth}
 \includegraphics[width=\textwidth,height=2cm]{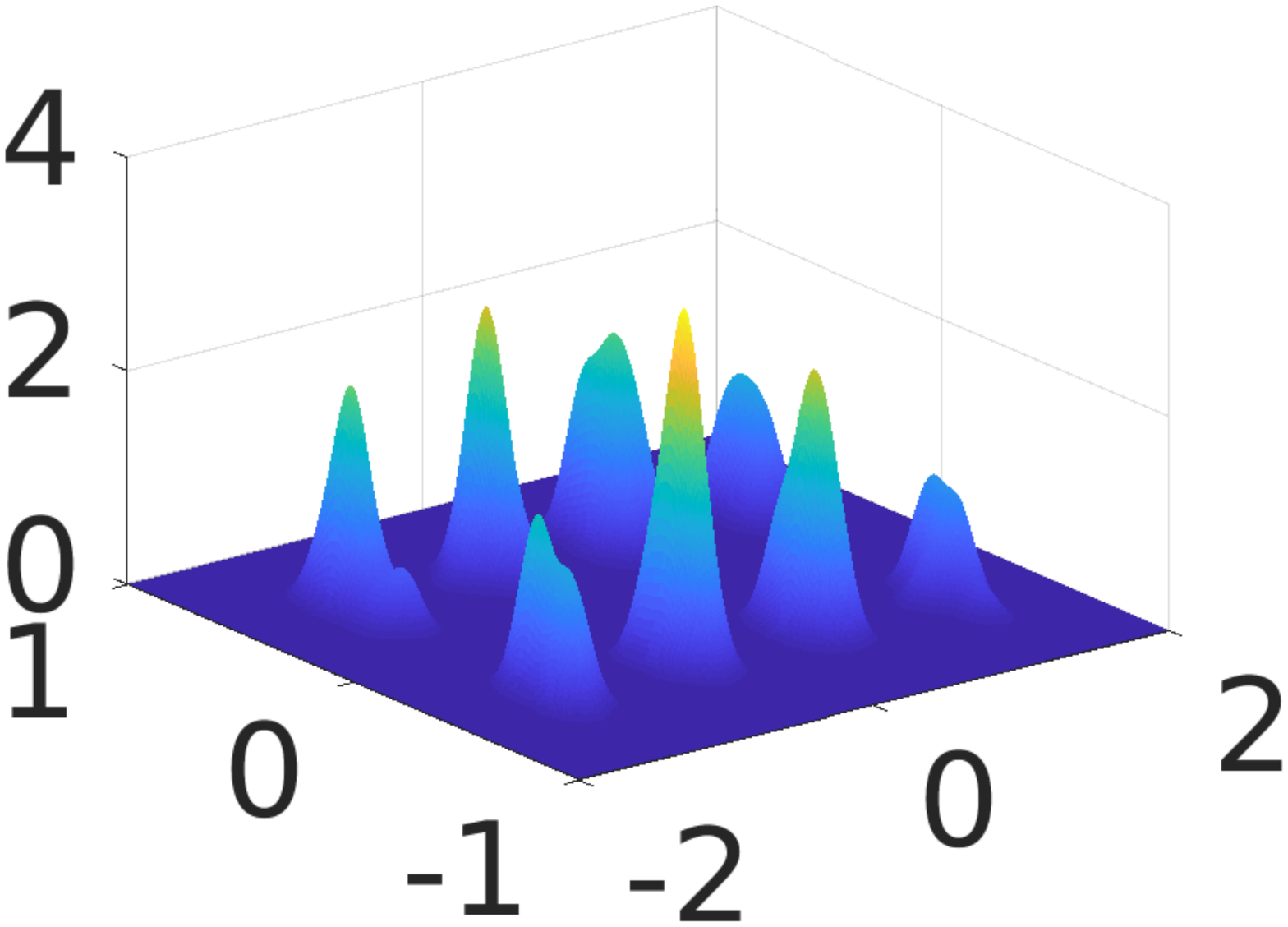}
 \caption{Imbalanced}\label{fig:imbalance}
 \end{subfigure}
 \hfill
 \begin{subfigure}[t]{0.28\linewidth}
 \includegraphics[width=\textwidth,height=2cm]{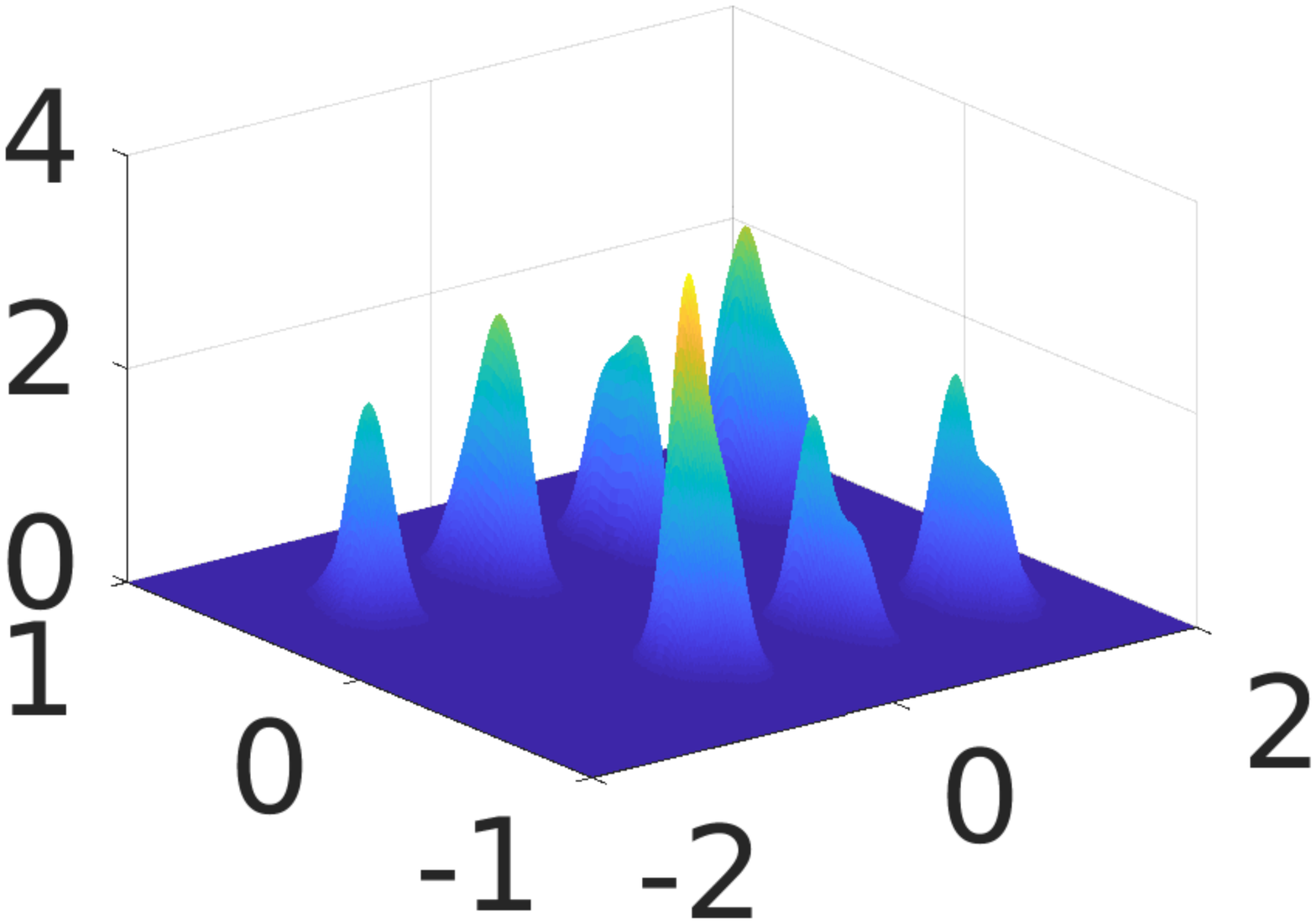}
 \caption{Missing}\label{fig:missing}
 \end{subfigure}
 \vsa
\caption{\footnotesize Existing modrec algorithms require uniform representation of constellation symbols as illustrated in the left-most figure. Biases due to imbalanced (middle) or missing symbols (right) leads to poor modrec.}\label{fig:bias}
\end{minipage}
\end{figure}

\noindent{\bf 2) Effects of bias in instance samples.} 
Prior work assumes that each symbol from a modulation's constellation is uniformly represented in a spectrum scan (e.g. Fig.~\ref{fig:uniform}). However, real world spectrum sensing and occupancy may introduce various biases in symbol representation. For example, scan intermittency may result in insufficient amount of \iq samples, which in turn may lead to non-uniform symbol representation. Biases are also possible due to malicious transmitters, which purposefully obfuscate the constellation symbols to deceive modrec algorithms~\cite{rahbari2016full}. Biases both due to small number of samples (Fig.~\ref{fig:imbalance}) or missing symbols (Fig.~\ref{fig:missing}) affect the overall constellation, and similar to partial scans, have a negative impact on the discriminative power of global features. 

\begin{figure}
\begin{subfigure}[t]{0.495\linewidth}
    \centering
    \includegraphics[width=\linewidth]{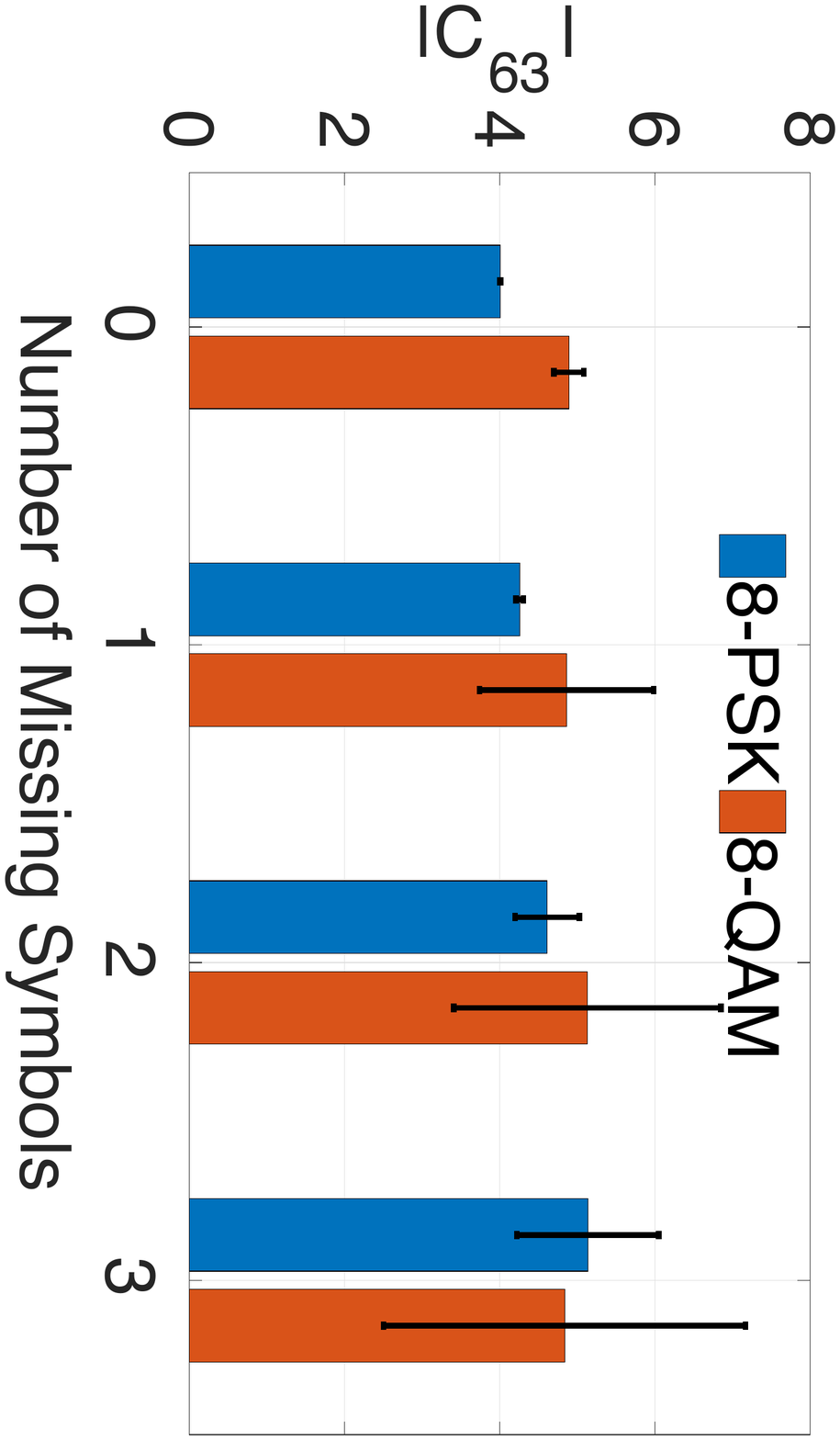}
    \caption{\footnotesize $C_{63}$ (missing symbols)}
    \label{fig:c63bias}
\end{subfigure}
\begin{subfigure}[t]{0.49\linewidth}
    \centering
    \includegraphics[width=\linewidth]{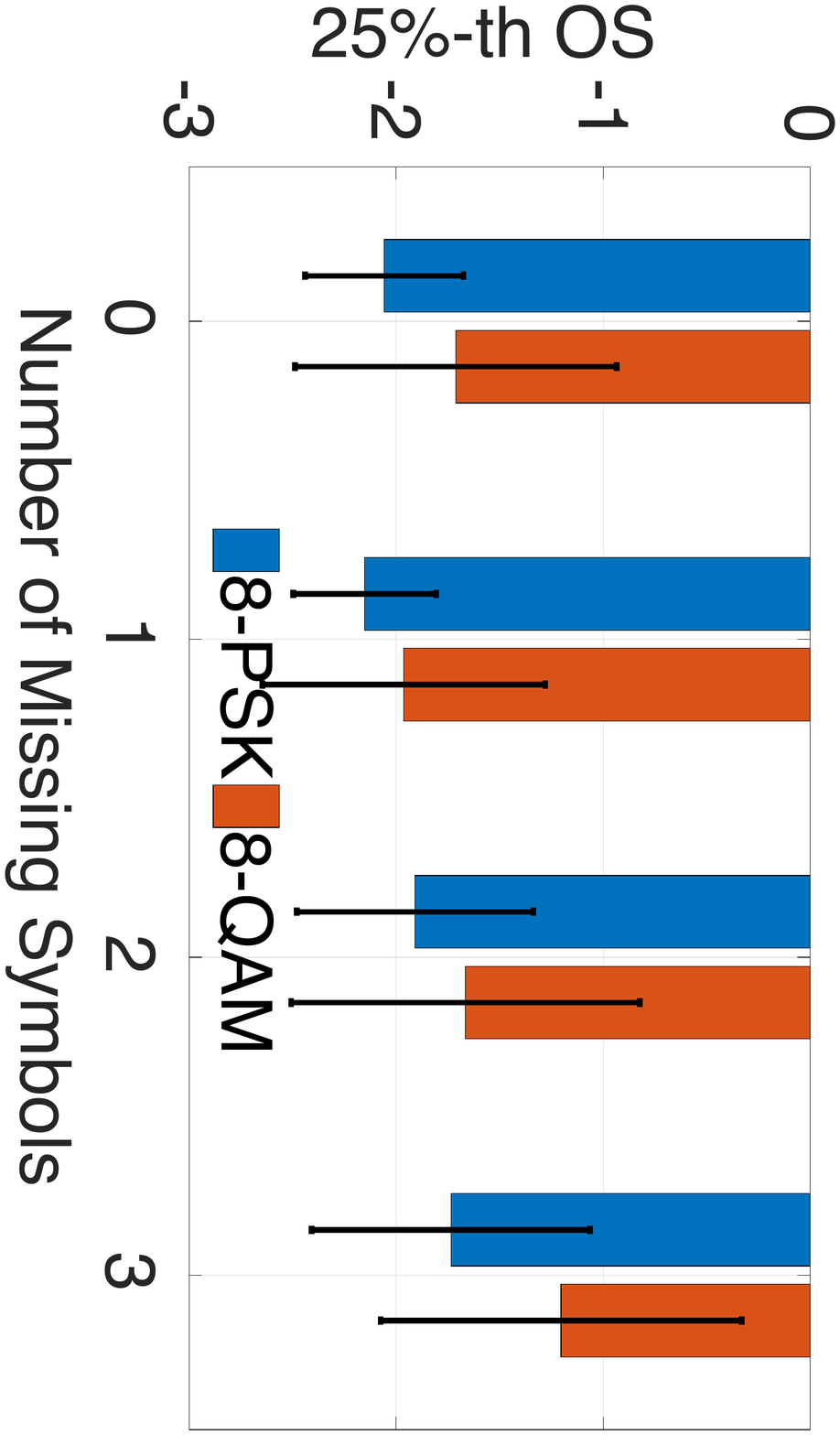}
    \caption{\footnotesize $25\%$ OS (missing symbols)}
    \label{fig:osbias}
\end{subfigure}
\vsa
\caption{\footnotesize Effects of constellation biases on the discriminative power of global features. Y-axis presents mean and standard deviations over $100$ instances of (\subref{fig:c63bias}) $C_{63}$ HOC and (\subref{fig:osbias}) the $25\%-$ile OS as a function of the number of missing symbols.}
\label{fig:global-stability-bias}
\end{figure}

We study these effects in Fig.~\ref{fig:global-stability-bias}. Fig.~\ref{fig:c63bias} and \ref{fig:osbias} show the behavior of $C_{63}$ and the $25\%$ OS for \texttt{8-PSK} and \texttt{8-QAM} with increasing number of randomly missing symbols. The respective feature values converge between modulation types, while their variance increases drastically with increasing number of missing symbols. Once again, this behavior suggests a deteriorating discriminative power of global features with missing symbols, further evaluated in \S\ref{sec:evaluation}.

\noindent{\bf 3) Effects of constellation rotation.} Global OS features assume $0^\circ$ rotation of the modulation's constellation~\cite{han2017low}, i.e. prior knowledge of the transmitter's technology. This may not be available when sensing arbitrary agile transmitters in the wild, once again negatively affecting the performance of global feature-base modrec (details in \S\ref{sec:evaluation}).

\section{\ourmeth Methodology} \label{sec:method}
Constellation biases, scan partiality, constellation rotation and increased noise levels all distort the global statistical properties of \iq instances, and thus, deteriorate the efficiency of corresponding modrec approaches. At the same time, these challenges are ubiquitous when the problem is considered in realistic settings. To improve the robustness of modrec techniques, we propose to capture information contained in the local ordering of \iq samples. The resulting local features are robust to imperfections due to real-world sensing, and when combined with global HOC features, enable high-accuracy modrec, exhibiting superior performance in both simulated and real-world scans. In what follows, we first present and overview of \ourmeth. We then present the intuition behind our proposed local features and describe the methodology for their extraction. Finally, we detail how these features can be employed in a classification framework for efficient modrec.

\vspace{-.4cm}
\subsection{Overview of \ourmeth}

\begin{figure*}[t]
\begin{minipage}{\linewidth}
\centering
\begin{subfigure}[b]{\textwidth}
\includegraphics[width=\textwidth]{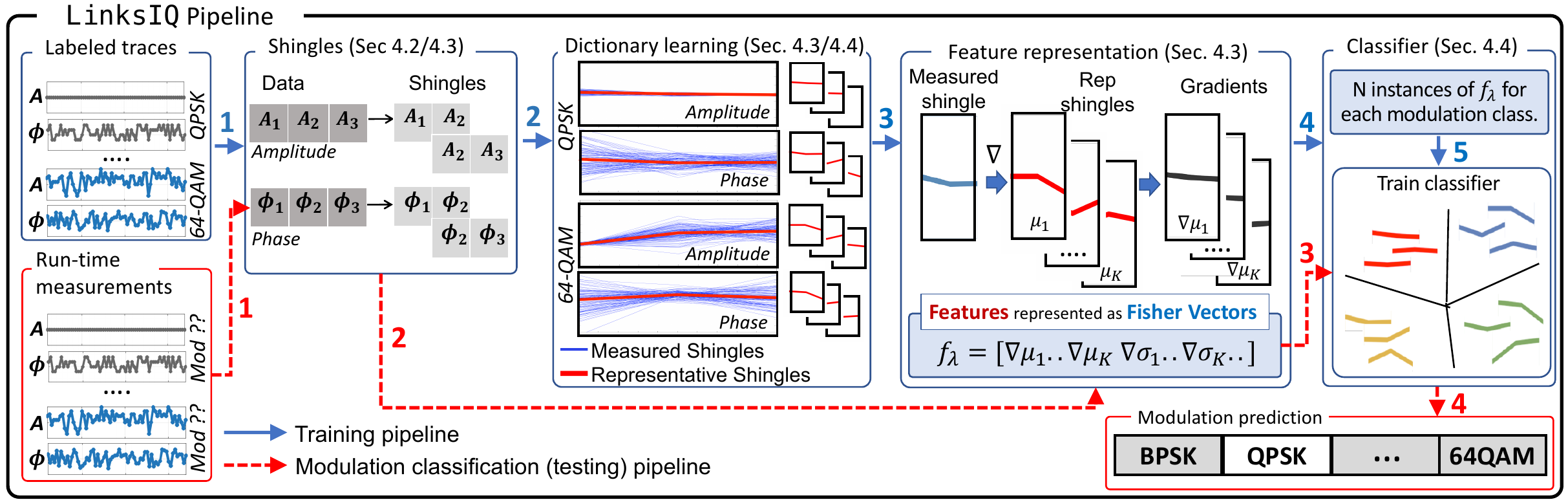}
\end{subfigure}
\caption{\footnotesize \ourmeth pipeline. The blue path traces steps involved in training, whereas the red path represents modulation classification (i.e. testing). }
\label{fig:system_pipeline}
\end{minipage}
\end{figure*}

Fig.~\ref{fig:system_pipeline} presents the operational pipeline of \ourmeth. The process consists of a \textit{training} and \textit{testing} (i.e. modulation classification) phase. Blue solid lines trace the training steps, whereas red dashed lines trace the classification steps. In training, we begin with the collection of a labeled training dataset, whereby spectrum traces are collected from transmissions with previously-known modulations. In step 1, we decompose the measured amplitude and phase time-series into shingles, as discussed in \S\ref{sec:intuition-local} and \S\ref{sec:dictionary1}. Following the shingle extraction, we learn a dictionary of representative shingles for each modulation, as detailed in \S\ref{sec:dictionary1} and \S\ref{sec:dictionary}. We then create the feature representation for each measured instance as described in \S\ref{sec:dictionary1}. Finally, we train an SVM classifier, as per \S\ref{sec:dictionary}.

The runtime classification, following red arrows in Fig.~\ref{fig:system_pipeline}, begins with the collection of unlabeled spectrum traces. Next, we compute the set of amplitude and phase shingles in the data. Then, in step 2, we find the feature representation of the measured shingles with respect to the representative shingles found in the dictionary learning phase. Finally, we employ the pre-trained classifier to determine what is the modulation of the measured signal. In what follows, we detail each of the steps in our pipeline.

\vspace{-.4cm}
\subsection{\iq sample sequences as a classification feature}
\label{sec:intuition-local}
While the order and relationships of individual \iq samples within an instance $x$ has not been considered in the modrec literature, we postulate that it carries important information, which is better preserved in realistic settings and can be used to improve modrec accuracy. This intuition is inspired by the tremendous success of local features extracted from images in computer vision and particularly employed for natural image classification~\cite{sanchez2013image,zhang2007local}. In our case, we treat an instance $x$ as a 1-dimensional signal as opposed to the typical 2D setting arising in computer vision.

\begin{figure}
    \begin{minipage}{.15\textwidth}
    \centering
    \includegraphics[width=.7\textwidth,height=3.2cm]{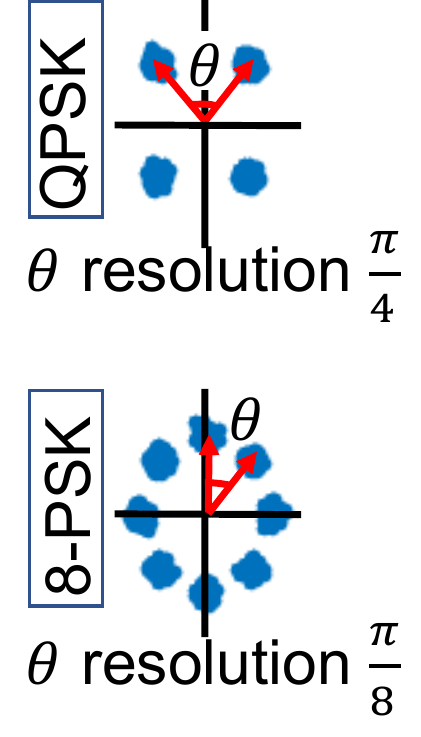}\vsa\vspace*{0.1cm}
    \caption{\footnotesize{Phase transitions; $\theta$=$\pi/8$ does not occur in QPSK.}}
    \label{fig:phasepattern}
    \end{minipage}
    \hfill \hfill
    \begin{minipage}{.34\textwidth}
    \vsc
    \centering
    \begin{subfigure}[b]{0.44\textwidth}
        \includegraphics[width=\textwidth]{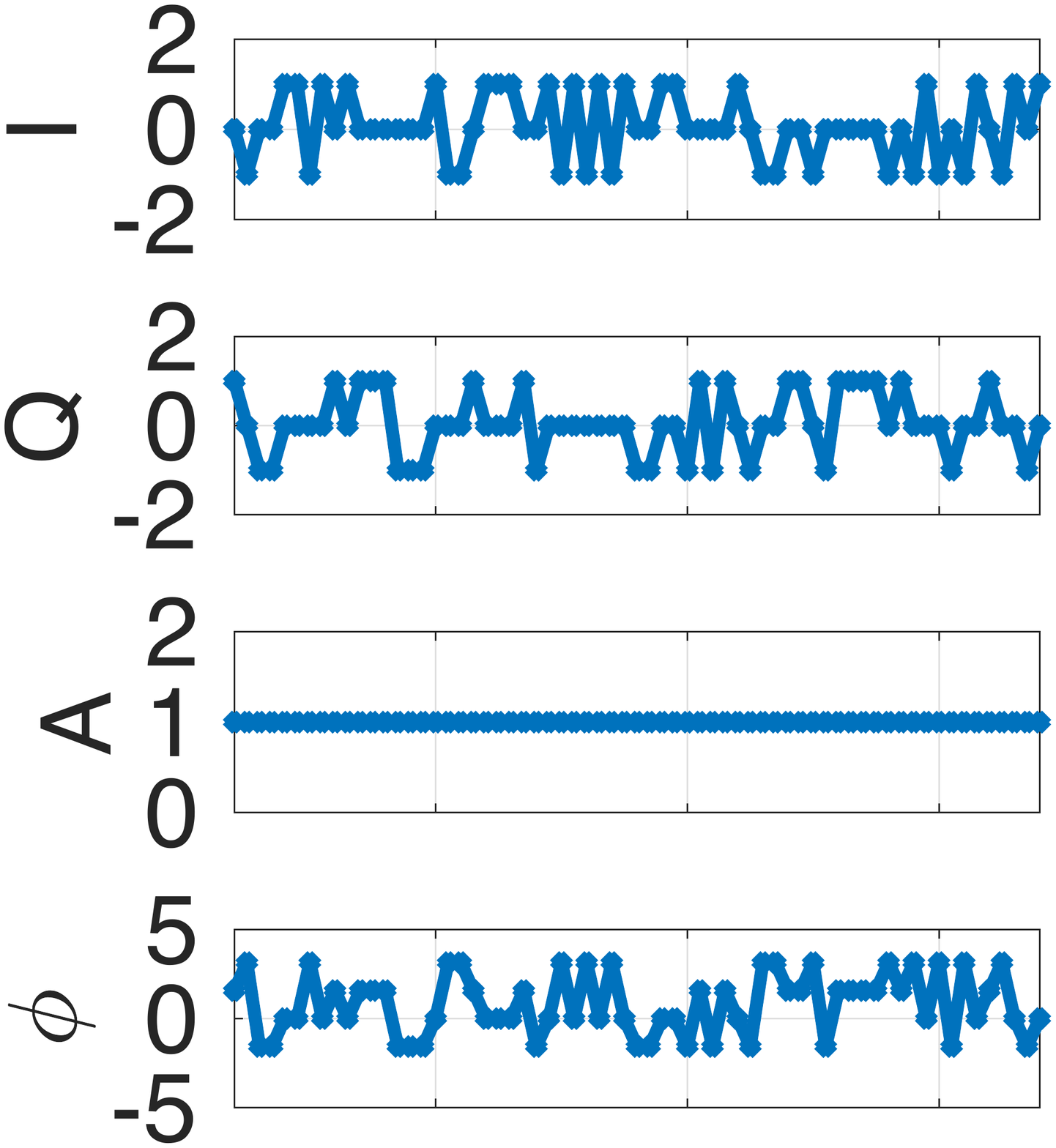}\vsc
        \caption{\texttt{QPSK}}
    \end{subfigure}
    ~ 
    \begin{subfigure}[b]{0.47\textwidth}
        \includegraphics[width=\textwidth]{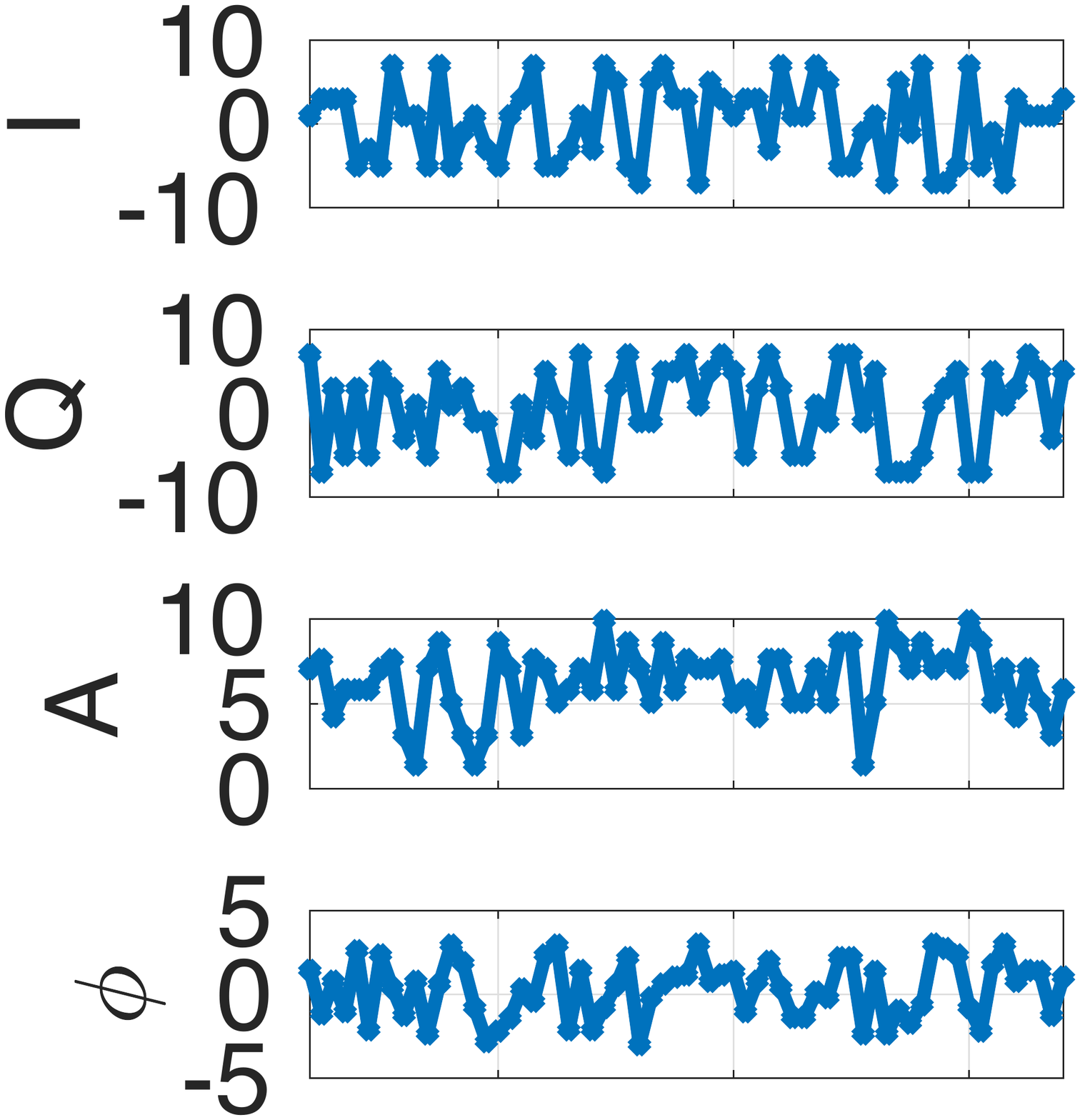}\vsc
        \caption{\texttt{64-QAM}}
    \end{subfigure}
    \caption{\footnotesize{LP of Polar and Cartesian representation of \texttt{IQ} samples over time (x-axis). Both magnitude and shape of the LPs vary across modulations.}}\label{fig:lpintuit}
    \end{minipage}
\end{figure}

To understand the intuition behind our approach, let us consider Fig.~\ref{fig:phasepattern} which presents the constellations of \texttt{QPSK} and \texttt{8-PSK} modulations. Intuitively, an instance $x$ comprised of \iq samples timeseries is a trajectory of transitions between the constellation points. The distribution of angular distances (angle changes) between consecutive transitions arising from different modulations varies due to the varying inter-cluster distances in their constellations. For example, transitions in \texttt{8-PSK} will be centered around multiples of $\pi/8$, while those in \texttt{QPSK} around multiples of $\pi/4$.
Fig.~\ref{fig:lpintuit} depicts a segment of the \iq sequences arising from \texttt{QPSK} and \texttt{64-QAM}. We plot separately the \texttt{I} and \texttt{Q} components as well as the corresponding amplitude $A$ and phase $\phi$ sequences. Qualitatively, it is evident that high-order modulations (i.e. \texttt{64-QAM}) exhibit bigger variation compared to their low-order counterparts. Furthermore, the phase transitions of low-order modulations are sharp from sample to sample, while these transitions are smoother with higher-order modulations.

To capture the sequential information encoded in \iq sample subsequences, we focus on intervals of their time domain and propose to extract modulation-specific transition signatures. We demonstrate that, such signatures are more robust to noise, sample bias, constellation rotation and partial transmitter overlap than global features alone, and thus can be employed to improve modrec accuracy.

\vspace{-.4cm}
\subsection{Learning local sequential features}\label{sec:dictionary1}
\label{sec:patch_extraction}
Let $x_A,x_\phi\in\mathbb{R}^n$ denote the real-valued sequences of observed amplitude and phase values in an instance $x$. We employ the same framework to extract local features from each of those sequences separately, as they are advantageous for different modulation types. For example, the amplitude sequence $x_A$ will be discriminative for amplitude-related modulation properties (e.g. \texttt{PSK} v.s. \texttt{QAM} families), while the phase sequence $x_\phi$ will be useful to differentiate phase-related properties (e.g. \texttt{QPSK} v.s. \texttt{8-PSK}). The same framework can be applied to the sequences of \texttt{I} and \texttt{Q} components, however, in our experimental evaluation they did not offer additional discriminative power. In what follows, we will simplify the notation by denoting $x=[x_1,x_2,\dots, x_n]$ as either of the real-valued sequences $x_A$ or $x_\phi$.

We adopt a generative framework to model a sequence $x$ in terms of all of its subsequences of length $l$, to which we will refer as \emph{shingles}. Specifically, let $x_i^l$ denote a shingle starting at position $i$ of length $l$. An instance $x$ of length $n$ has a total of $n-l+1$ such shingles. Our key assumption is that observed instance shingles are generated from some parametric generating distribution $p_\lambda$ parametrized by a set of parameters $\lambda$. This representation is similar to n-gram based models for text~\cite{cavnar1994n} and patch-based representations for images~\cite{sanchez2013image}.

We adopt a \emph{Gaussian Mixture Model (GMM)} as the generating distribution $p_\lambda$, which is a typical choice in patch-based representation of images~\cite{sanchez2013image}. A $K$-component GMM is fully specified by $\lambda=\{w_k,\mu_k,\Sigma_k\}, k=1 \dots K$, where  $w_k\geq0$ is the non-negative mixing weight of the $k$-th component and $\mu_k$ and $\Sigma_k$ are its mean vector and covariance matrix respectively. We further disregard mixed covariance terms for shingles and instead work with a variance vector $\sigma_k^2$ (i.e. we assume a diagonal covariance matrix). This assumption is justified in our case as consecutive constellation symbols within shingles are determined by the encoded data and we do not place any assumptions on their sequence. Note that the shingle size $l$ determines the dimensions of $\mu_k$ and $\sigma_k^2$.

We adopt the \emph{Fisher Kernel (FK)} representation which defines similarities between shingles in terms of dot products of their \emph{Fisher Vectors (FVs)}~\cite{jaakkola1999exploiting}. Formally, a FV $f_\lambda(x_i^l)$ representing shingle $x_i^l$ is defined as:

\begin{equation}
f_\lambda(x_i^l) = L_\lambda \nabla_\lambda \log{ p_\lambda(x_i^l)},
\label{eq:fisher_vector}
\end{equation}
where $\nabla_\lambda \log{ p_\lambda(x_i^l|\lambda)}$ is the gradient of the log-likelihood of the observed shingle $x_i^l$ being generated by $p_\lambda$, where the gradient is evaluated at $x_i^l$; and $L_\lambda$ is the square root of the inverse of the \emph{Fisher Information Matrix (FIM)}.
$L_\lambda$ normalizes the dynamic range of gradient vectors similar to its use in~\cite{jaakkola1999exploiting}. 

We obtain the local feature representation $f_\lambda(x)$ of the whole instance $x$, given a GMM model $p_\lambda$, as the average Fisher Vector of all observed shingles within the instance:
\begin{equation}
f_\lambda(x) = \frac{1}{n-l+1}\sum_{i=1}^{n-l+1}f_\lambda(x_i^l).
\end{equation}
In other words, the instance FV is the average of the normalized gradient statistics of all involved shingles, where $L_\lambda$ is treated as a normalization factor. We apply the same transformation to both the amplitude $x_A$ and phase $x_\phi$ sequences and concatenate the resulting FVs.
In what follows, we discuss how to derive the normalization $L_\lambda$ and gradient statistics $\nabla_\lambda \log{ p_\lambda(x_i^l|\lambda)}$ for individual shingle FVs.

The likelihood $p_\lambda(x_i^l)$ in GMM is defined as the average weighted likelihood of the shingle $x_i^l$ arising from the individual Gaussian components:
\begin{equation}
p_\lambda(x_i^l)=\sum_{k=1}^Kw_kp_k(x_i^l),
\end{equation}
where $p_k(x_i^l)$ is the pdf of the $k$-th $l$-variate Gaussian component in the GMM. To ensure that $p_\lambda(x_i^l)$ is a valid probability distribution the weights need to be all non-negative and sum to $1$, i.e. $\sum_{k=1}^Kw_k=1$. We use the gradient statistics with respect to the mean $\mu_k$ and variance $\sigma_k$ vectors of each component resulting in the following component-wise $L_\lambda$-normalized gradients:
\begin{equation}
f_{\mu_k}(x)=\frac{\nabla_{\mu_k}\log{ p_k(x)}}{\sqrt{w_k}}= \frac{\gamma_k(x)}{\sqrt{w_k}}\Big[\frac{x-\mu_k}{\sigma_k^2}\Big]
\label{eq:grad-mu}
\end{equation}
\begin{equation}
f_{\sigma_k}(x)=\frac{\nabla_{\sigma_k}\log{ p_k(x)}}{\sqrt{w_k}}= \frac{\gamma_k(x)}{\sqrt{w_k}}\Big[\frac{(x-\mu_k)^2}{\sigma_k^3}-\frac{1}{\sigma_k}\Big],
\label{eq:grad-sigma}
\end{equation}
where the $\gamma_k(x)$ is the soft assignment (posterior probability) of the shingle to component $k$ defined as:
\begin{equation}
\gamma_k(x)=\frac{w_kp_k(x)}{\sum_{i=1}^{K}w_ip_i(x)},
\end{equation}
and where exponentiation and division operations involving vectors $x$, $\mu_k$ and $\sigma_k$  in Eqs.~\ref{eq:grad-mu},~\ref{eq:grad-sigma} are element-wise operations (recall that they are $l$-dimensional vectors). Note that we do not consider the gradient statistic with respect to $w_k$ in our FV representation, arriving at a $(4lK)$-dimensional vector, representing $2$ series (amplitude and phase), maintaining shingle-length (i.e. $l$-dimensional) gradient statistics (both mean and variance) for each of the $K$ components of the GMM. We omit a gradient statistic with respect to the component weights $w_k$ which could be interpreted as prior component probabilities, as they require more data to robustly estimate (GMM model estimation is discussed next), than their variance and mean vectors. Investigation of whether these additional statistics boost the performance might be a fruitful further direction.

\begin{figure}[t]
\centering
    \centering
    \includegraphics[width=0.7\linewidth]{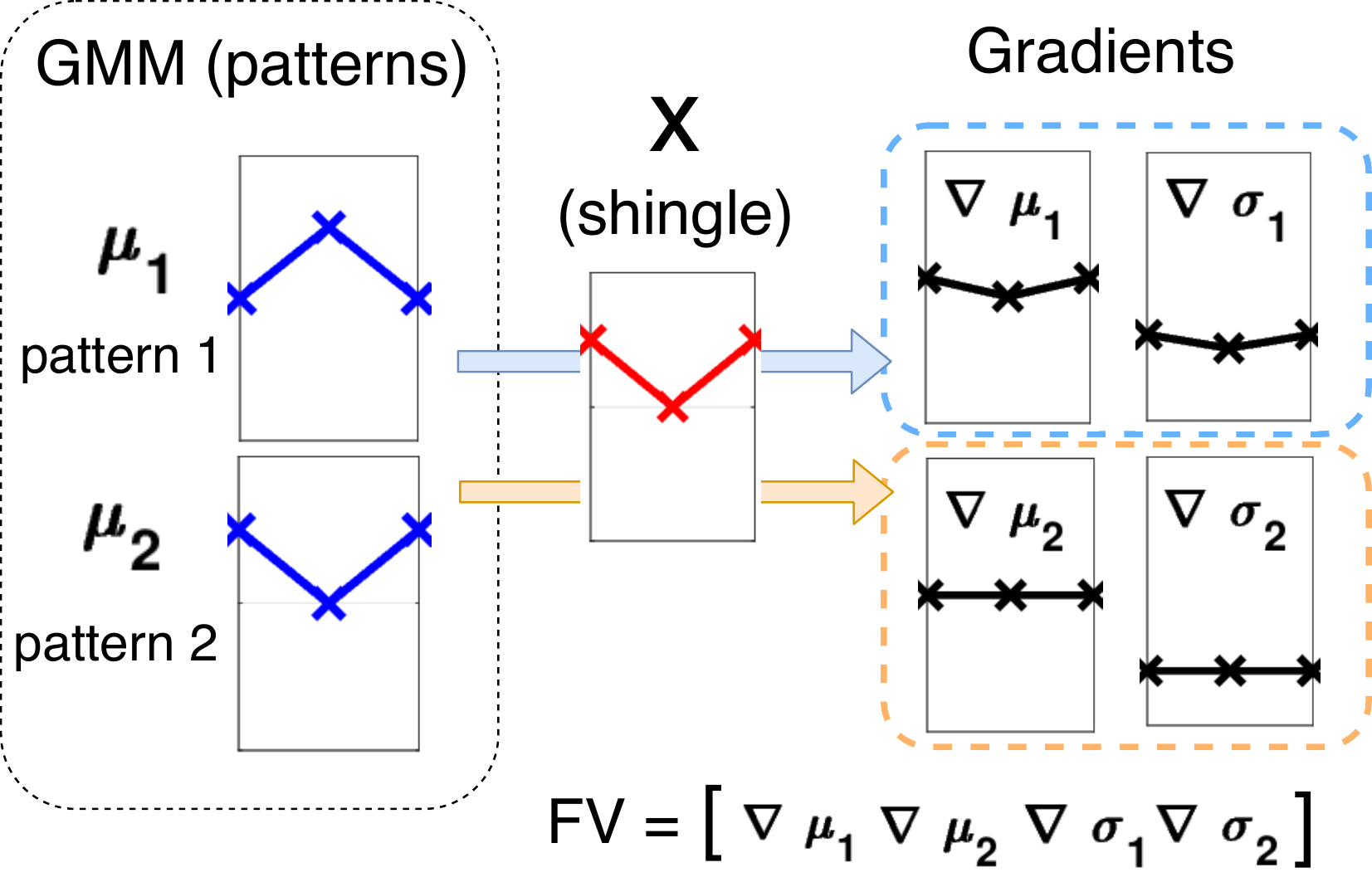}
    \vsa
\caption{\footnotesize Example of local features computed for $l=3$-dimensional shingle $x$ and $2$-component GMM with $\mu_2$ agreeing ``better'' with the shingle than $\mu_1$ ($\sigma_{1,2}=\mathbb{1}$, $w_{1,2}=0.5$). The gradient statistics for each of the components are shown on the right. The mean gradient statistics of the ``better''-agreeing component $\nabla_{\mu_1}$ is closer to $\mathbb{0}$ (expected as per Eq.~\ref{eq:grad-mu}) and its $\nabla_{\sigma_1}$ is closer to $-\mathbb{1}$ (as per Eq.~\ref{eq:grad-sigma}).}
\vspace{-.5cm}
\label{fig:shingle-example}
\end{figure}

An illustrative example of the gradient evaluation for a shingle $x$ in a two-component mixture model with unit variance vectors is presented in Fig.~\ref{fig:shingle-example}. The well-agreeing GMM components result in close-to optimal corresponding gradient statistics. The final FV is composed of concatenating $[\nabla_{\mu_1} \nabla_{\mu_2} \nabla_{\sigma_1} \nabla_{\sigma_2}]$ (normalization by $1/\sqrt{1/2}$ omitted).

It is worth noting that, while resorting to a kernel method for representation of our local features as opposed to working directly with component likelihoods $p_k(x)$, results in higher dimensional representation, it comes with the usual advantages. Namely, when the kernel is appropriately selected, it allows modelling non-linear data using simple linear classifiers. In addition, the specific FV kernel has been shown to perform very well in the natural images domain and typically requires small number of Gaussian components for good discriminative power, thus allowing good scalability~\cite{sanchez2013image}. We experimented with non-kernel local feature representations and did not find similar improvements over state-of-the-art global feature methods as the ones exhibited by the FV kernel representation.

\vspace{-.4cm}
\subsection{Model learning: GMM dictionary and classification}\label{sec:dictionary}

To enable modrec employing our local features, we need to first estimate a GMM model from shingle observations in actual instances and then train a modulation classifier based on the feature encoding of instances.

\noindent{\bf GMM dictionary learning.} The FV representation outlined in \S\ref{sec:patch_extraction} depends on a GMM generating distribution for shingles. Intuitively, we need to learn a ``dictionary'' of prototypical shingles, observed in instances across modulations and learn their component-wise mean $\mu_k$, variances $\sigma_k$ and relative weights $w_k$. Given a fixed dictionary size $K$ and a shingle length $l$, we learn a GMM based on a training data set $X$ containing instances of all modulation classes we aim to predict. Note, that since we do not use the class information $y$ associated with instances in $X$, our dictionary GMM learning is unsupervised. Supervised alternatives may allow even sparser discriminative representations for classification~\cite{mairal2009supervised}, however, we leave this direction for future exploration. To learn the GMM model from a training set $X$ we first extract shingles from the instances and use the seminal Expectation Maximization (EM) approach~\cite{bishop2006pattern}. Details about selecting the dictionary size $K$ and shingle length $l$ are discussed in \S\ref{sec:evaluation}.

\noindent{\bf Classification.} As we discuss earlier, the advantage of our Fisher Vector approach is that it captures non-linear information in its representation, and hence, simple classification techniques are expected to perform well. Thus, we adopt a simple linear SVM classifier with soft margin for our modrec task~\cite{cortes1995support}. We expect that other classification schemes may further improve the classification performance, but resort to a simple SVM in this work as our goal is to evaluate the utility of our local features and also employ a classifier which is typically employed by baseline global feature methods.

Our local feature scheme captures local transition information, however, we expect that the global sample distribution statistics may encode additional non-redundant information and thus consider classification schemes in which we concatenate the fisher vector $f_\lambda$ with the $7$ HOC features widely adopted in prior work. This combination is expected to ``lift'' the modrec performance of local features alone, particularly when the dictionary is learned on a rotated constellation w.r.t. that used in testing instances. We confirm this expectation empirically in \S\ref{sec:evaluation}.

\vspace{-.4cm}
\subsection{Algorithmic complexity}

Both the dictionary learning process and classifier training do not need to be repeated during actual modulation recognition, as long as they are performed on a training set that features instances from all target modulations. Thus, both processes can be thought of as ``offline'', i.e. they do not occur during actual modrec at work.

The complexity of modrec with our employed local features is the cost of encoding shingles from an instance $x$. Asymptotically, it depends on the dictionary and shingle sizes and and the number of samples instances $O(nlK)$, as there are $O(n)$ shingles in an instance and their gradient statistics of size $l$ need to be evaluated with respect to each of the $K$ GMM dictionary components. In practice we resort to short $l=3$ shingles and small dictionary size $K=50$ as they show optimal performance. Thus, assuming that $K$ and $l$ are constants relative to the number of samples $n$, the complexity of local patterns is linear $O(n)$ similar to that for computing HOC~\cite{swami2000hierarchical} and asymptotically better than OS (when using all samples)~\cite{han2017low} which require sorting the samples in $O(n\log n)$.

\vspace{-.4cm}

\section{Evaluation}\label{sec:evaluation}
We evaluate the robustness of our methodology with partial, biased and noisy scans in over-the-air and simulated settings. We begin by describing our implementation and data sets. In \S\ref{sec:bias}-\S\ref{sec:rotation} we evaluate the modrec performance of our method compared to state of the art HOC~\cite{swami2000hierarchical} and OS~\cite{han2017low}. For these results, we vary the classifier training, while using a universally-trained dictionary, as described in \S\ref{sec:data}. Unless otherwise noted, all accuracy results were obtained as an average from a 10-fold validation. 

\vspace{-.4cm}
\subsection{Implementation, data and parameters}\label{sec:data}

\noindent \textbf{Implementation.} Our method is implemented in MATLAB with all experiments executed on Ubuntu 14 machines. The Fisher Vector dictionary learning module is implemented using \cite{vedaldi2010vlfeat}. For classification, we adopt the SVM classifier model from MATLAB. We use one-versus-rest label coding to transform multi-classification to binary classification. We use the same classification approach across all compared features (i.e. HOC, OS, LP and LP+HOC).

\noindent \textbf{Data.} We use two datasets for our evaluation: one generated in a MATLAB simulation and one from a software-defined radio testbed. For our simulation we use MATLAB Communications System Toolbox to implement a transmitter and receiver connected by a AWGN channel. The transmitter is configured to use \texttt{QPSK}, \texttt{8-PSK}, \texttt{8-QAM}, \texttt{16-QAM} and \texttt{64-QAM}. We tune various blocks of our transmitter-receiver chain to generate the necessary datasets as follows. For \textit{partial scans}, we tune the low-pass filter at the receiver side by setting its cut-off to a fraction of the transmitter's bandwidth. For \textit{biased scans}, we purposely modify the input signal at the transmitter side to reduce or remove the occurrence of a given symbol. To control the \textit{noisiness} of the collected scan we tune the SNR level of the AWGN channel. Finally, to control the \textit{constellation rotation}, we modify the modulation block at the transmitter side. Our simulation-based evaluation is presented in \S\ref{sec:bias}-\S\ref{sec:rotation}. We also present results from partial scans from USRP-based transmissions, in a heterogeneous sensor testbed as detailed in \S\ref{sec:eval_real}. 

\noindent\textbf{Default parameters.} All performance results presented in \S\ref{sec:evaluation} were obtained with a single universal dictionary of patches trained at SNR 10dB, with no data bias, at $100\%$ transmitter overlap with mixed constellation rotation. The patch size, is set to 3 and the dictionary size to 50. A natural question is whether the dictionary learning parameterization (i.e. how we set the patch and dictionary size) and training data play role in our algorithm's performance. We explore this question in \S\ref{sec:dl} and show that the above universally-trained dictionary is feasible across all real-world settings. 

\vspace{-.3cm}
\subsection{Robustness to data bias}\label{sec:bias}

\begin{figure}[t]
\centering
\begin{subfigure}[t]{0.37\linewidth}
    \centering
    \includegraphics[width=\linewidth]{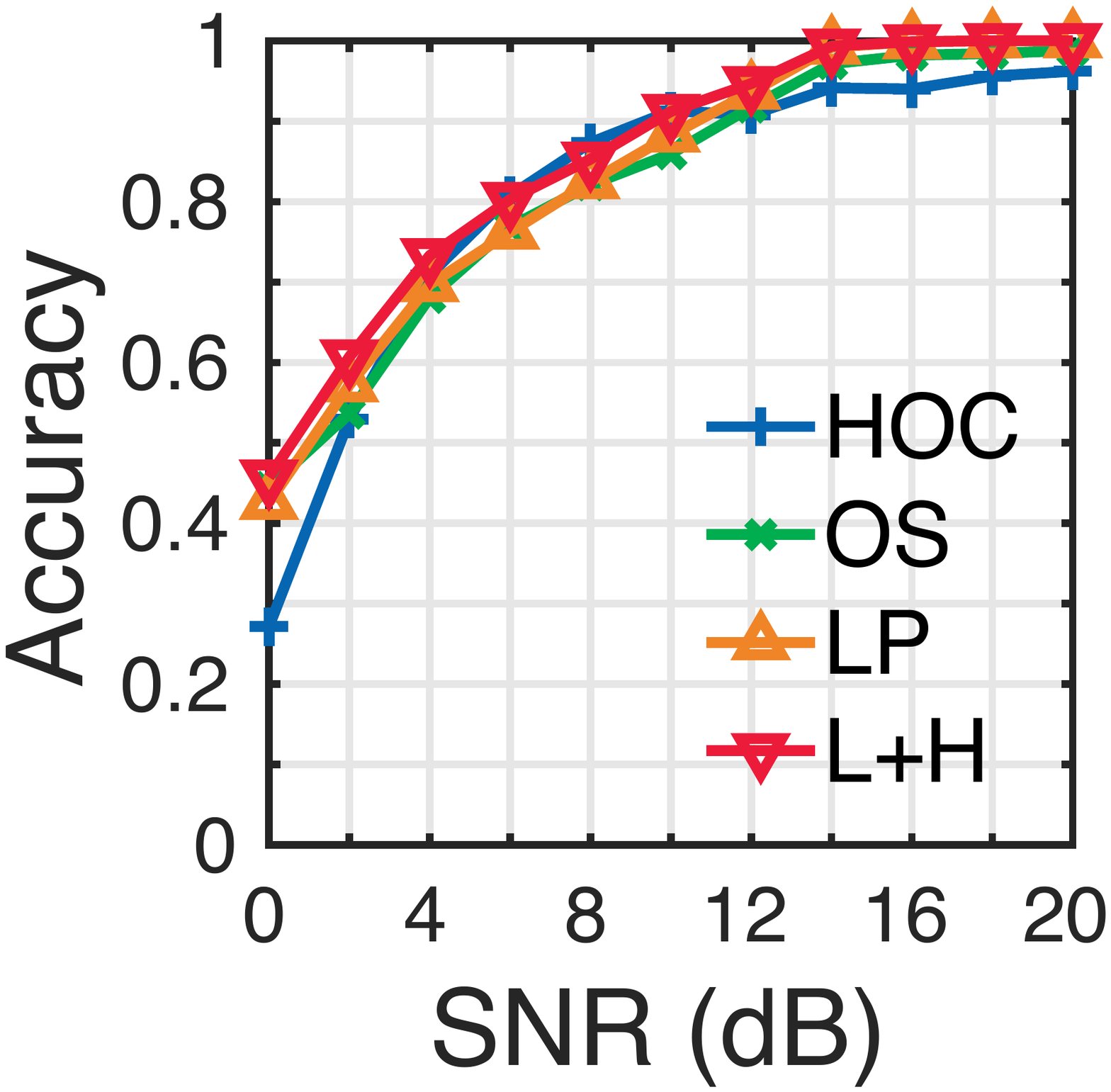}
    \caption{\footnotesize No missing symbols}
    \label{fig:bias0}
\end{subfigure}
\begin{subfigure}[t]{0.37\linewidth}
    \centering
    \includegraphics[width=\linewidth]{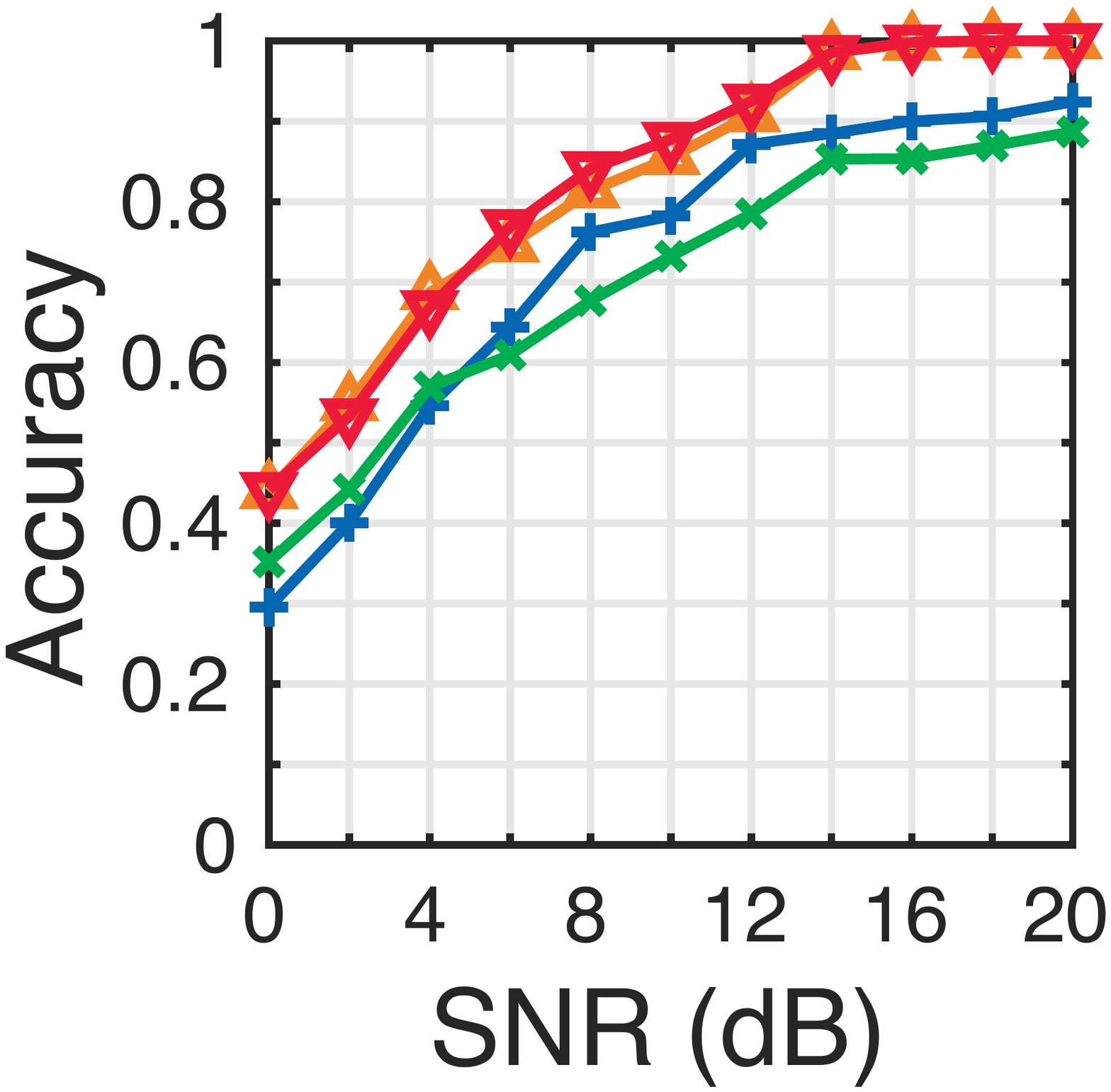}
    \caption{\footnotesize 1 missing symbol}
    \label{fig:bias1}
\end{subfigure}
\begin{subfigure}[t]{0.37\linewidth}
    \centering
    \includegraphics[width=\linewidth]{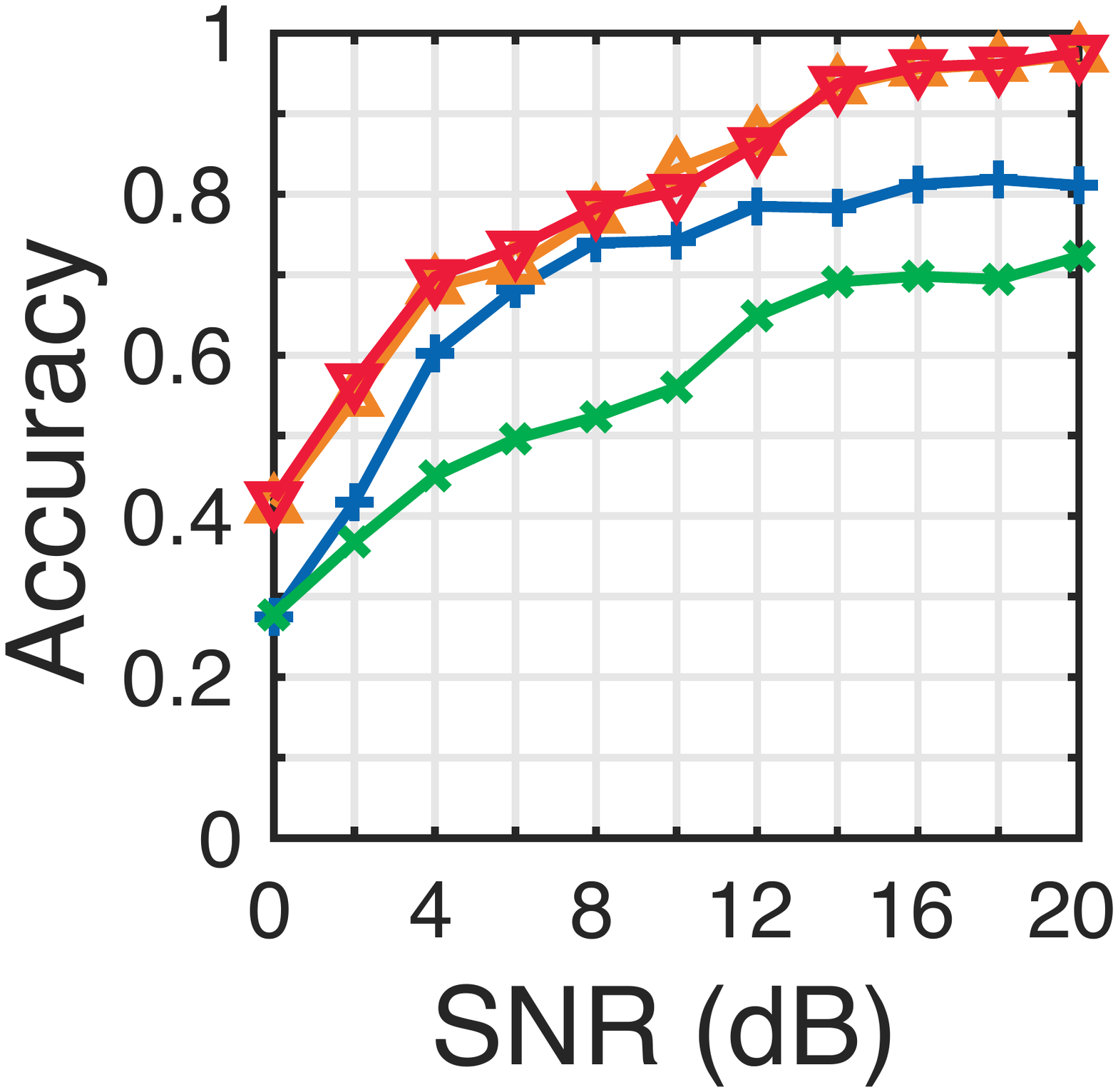}
    \caption{\footnotesize 2 missing symbols}
    \label{fig:bias2}
\end{subfigure}
\begin{subfigure}[t]{0.37\linewidth}
    \centering
    \includegraphics[width=\linewidth]{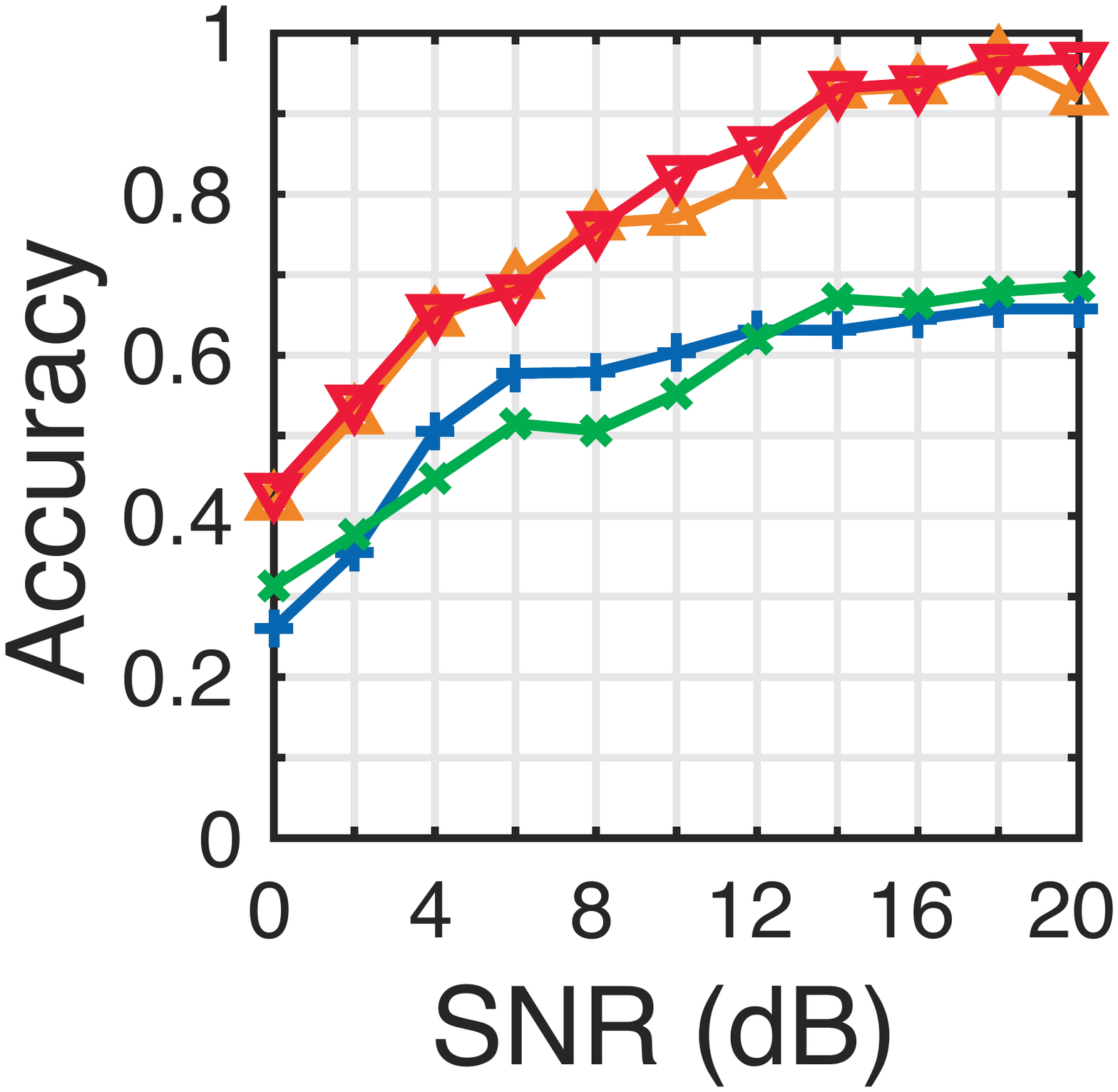}
    \caption{\footnotesize 3 missing symbols}
    \label{fig:bias3}
\end{subfigure}
\vspace{-.2cm}
\caption{\footnotesize{Performance with missing symbols. Classifier is SNR-aware and trained on unbiased data. Scans cover 100\% of the transmitter's bandwidth.}}
\label{fig:clf_accr_snr_missing_symbol}
\vspace{-.4cm}
\end{figure}

We begin by evaluating our method with data bias. All scans were collected at 100\% coverage of the transmitter bandwidth. We train the classifier on data with equal representation of all constellation symbols (i.e. unbiased data). A separate classifier was trained for each SNR level (i.e. classification is SNR-aware). We then test using data with purposely removed 1, 2 or 3 symbols. Fig.~\ref{fig:clf_accr_snr_missing_symbol} presents our results. For unbiased data (Fig.~\ref{fig:bias0}) all methods perform similarly. As bias is introduced, methods using global features deteriorate immediately even with SNR of 20dB. At 3 missing symbols global features can achieve a maximum of 69\% accuracy at 20dB, whereas our method maintains high accuracy of 98\%. Table~\ref{tab:breakdown} (left) shows a breakdown of performance of LP+HOC across modulations at SNR=10dB. For low-order modulations the accuracy is maximal and decreases as modulation order increases. These results demonstrate the potential of LP+HOC to successfully detect a transmitter's modulation in the face of data bias.

\vspace{-.3cm}
\subsection{Robustness to scan partiality}\label{sec:partial}

\begin{figure}[t]
\centering
\begin{subfigure}[t]{0.37\linewidth}
    \centering
    \includegraphics[width=\linewidth]{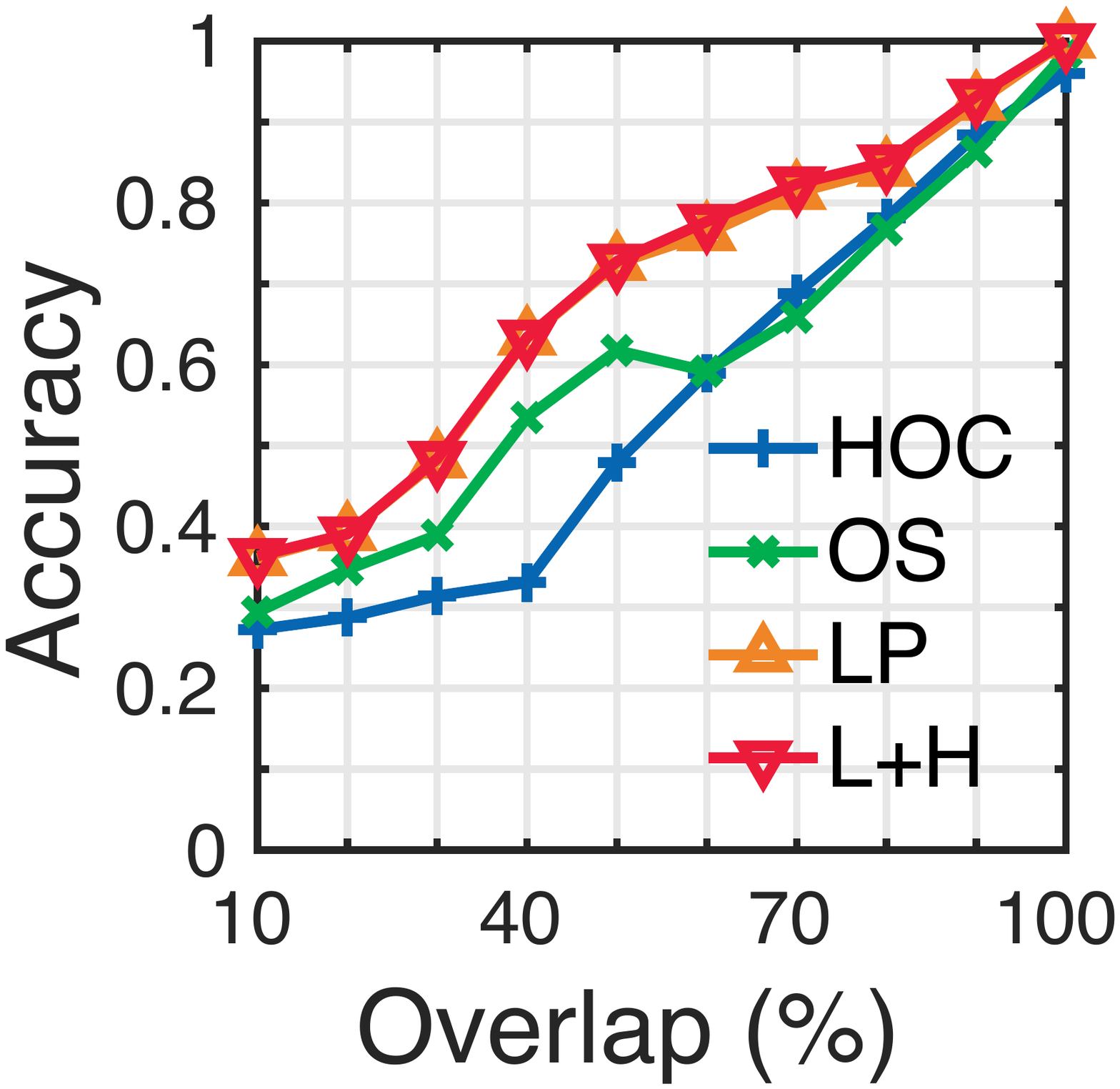}
    \caption{\footnotesize Overlap-aware; 20dB}\label{fig:partial_acc_20db}
\end{subfigure}
\begin{subfigure}[t]{0.37\linewidth}
    \centering
    \includegraphics[width=\linewidth]{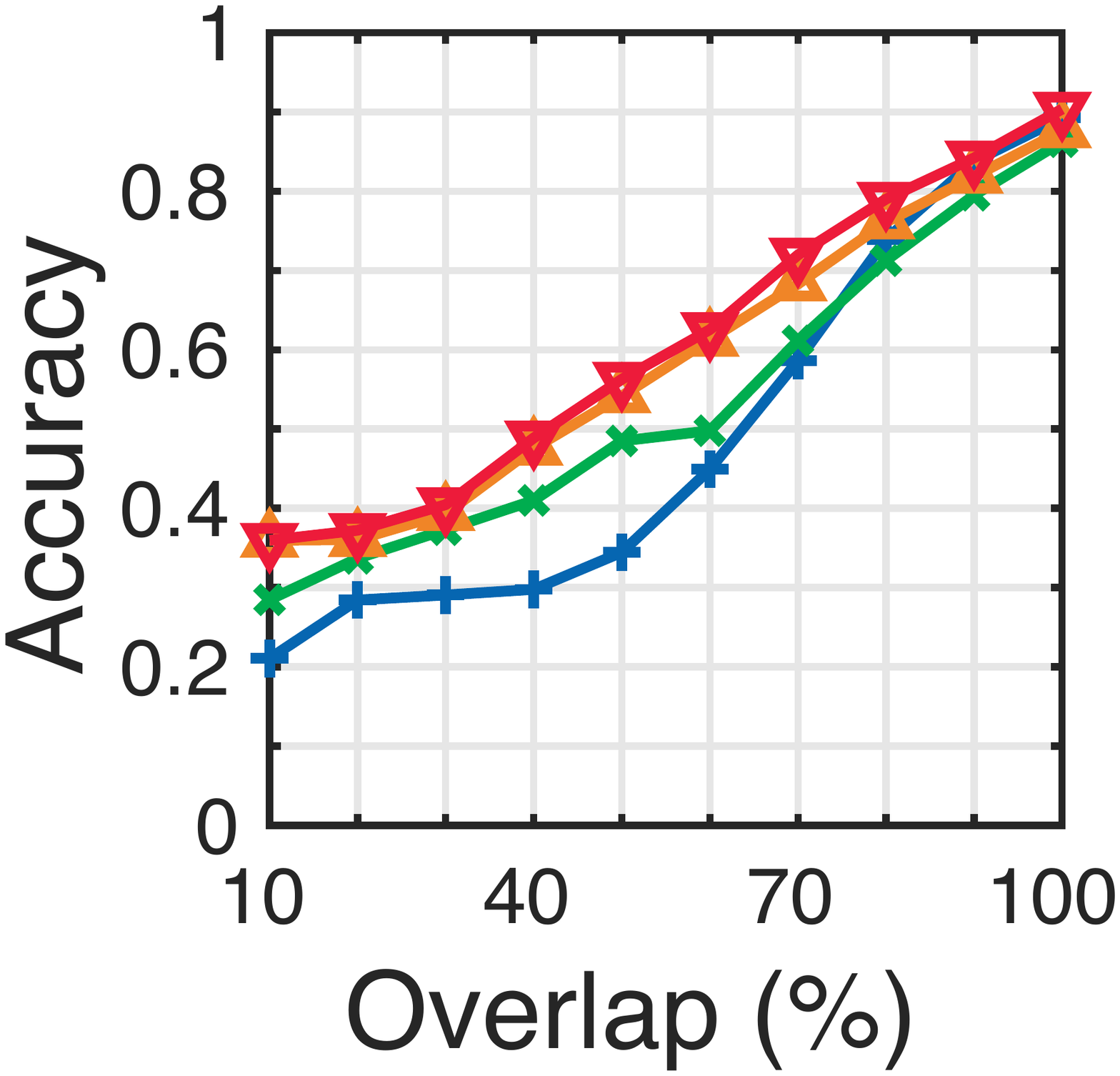}
    \caption{\footnotesize Overlap-aware; 10dB}\label{fig:partial_acc_10db}
\end{subfigure}
\begin{subfigure}[t]{0.37\linewidth}
    \centering
    \includegraphics[width=\linewidth]{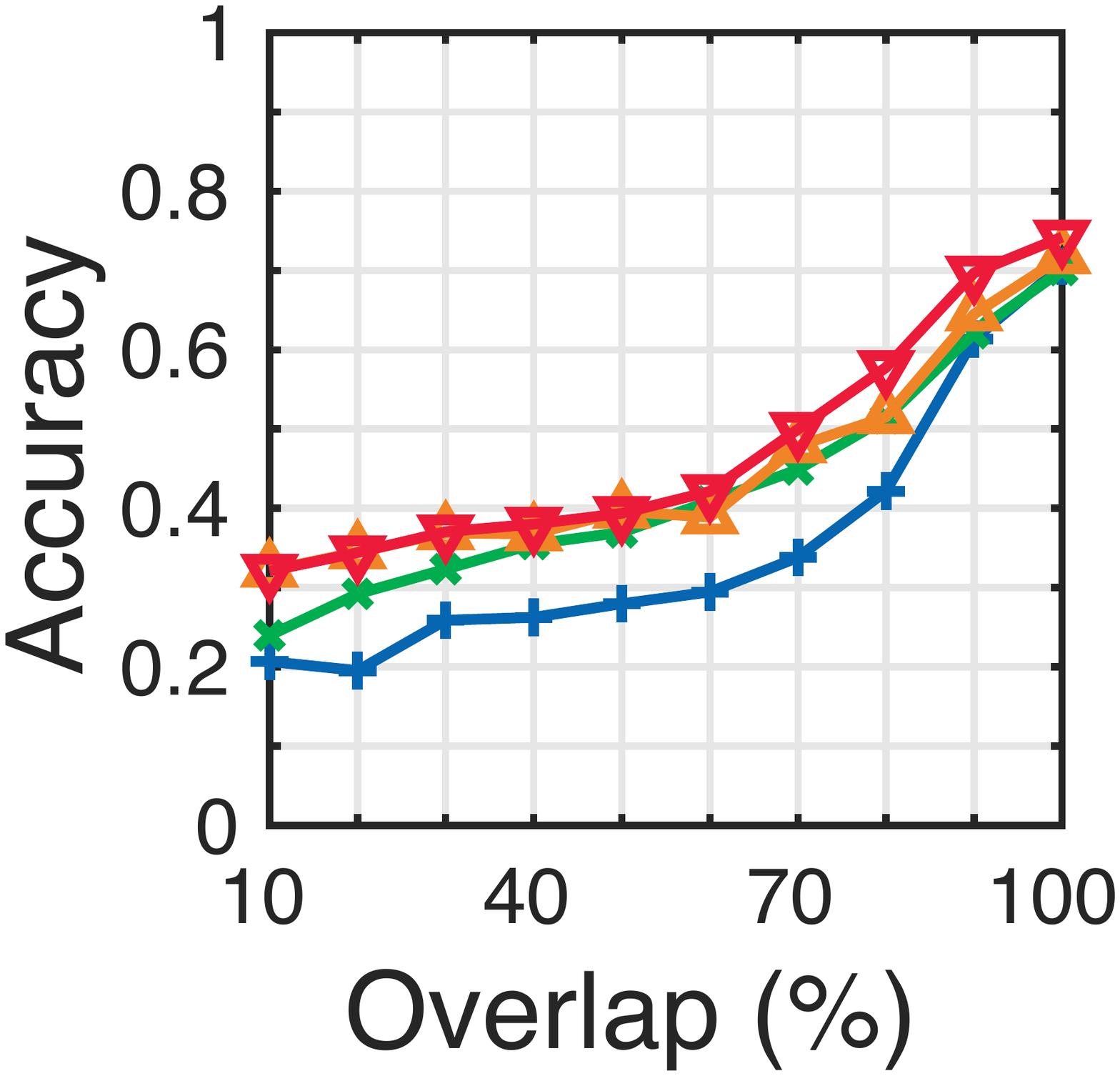}
    \caption{\footnotesize Overlap-aware; 4dB}\label{fig:partial_acc_4db}
\end{subfigure}
\begin{subfigure}[t]{0.37\linewidth}
    \centering
    \includegraphics[width=\linewidth]{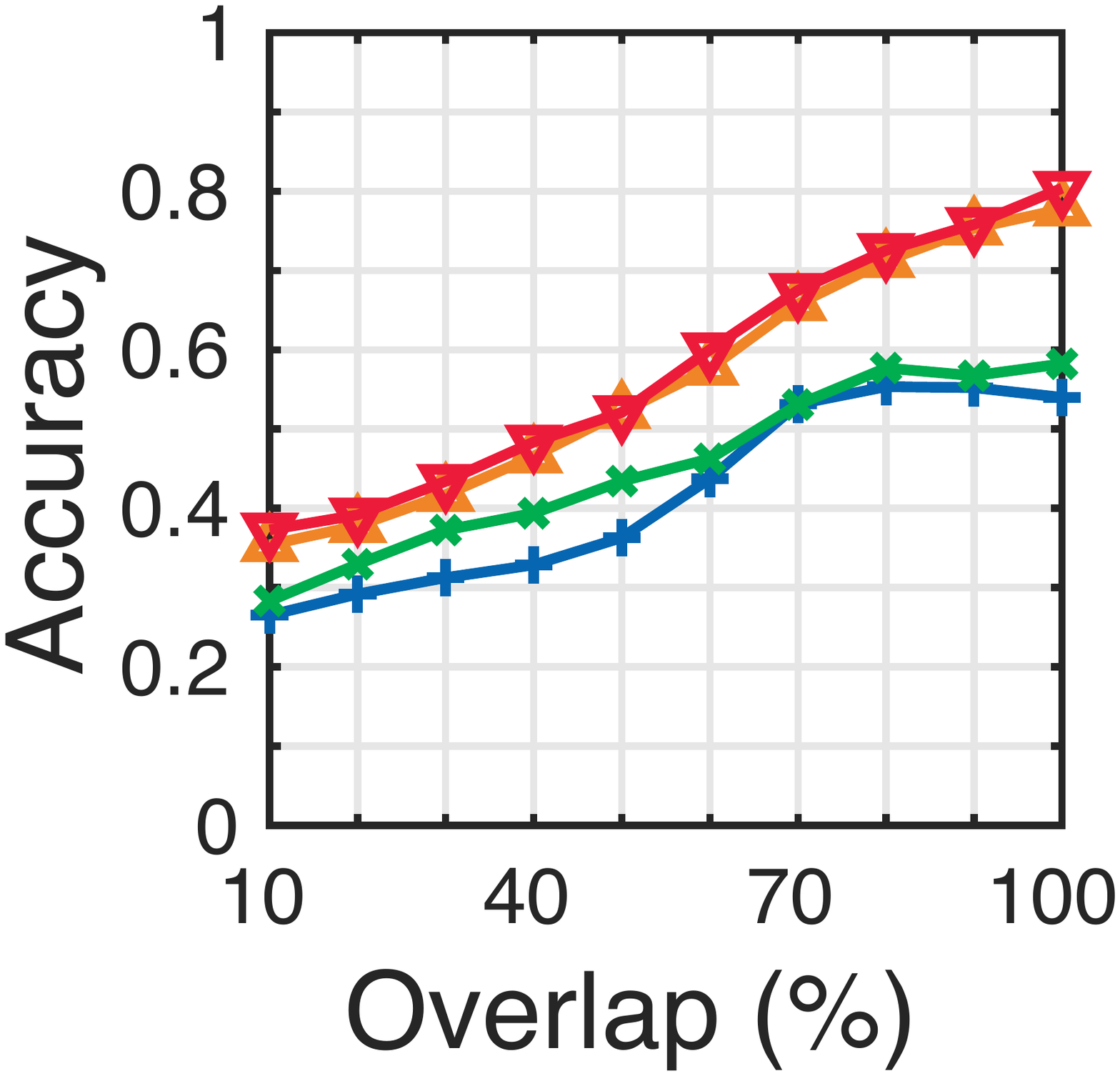}
    \caption{\footnotesize Overlap-blind; 10dB}\label{fig:partial_acc_blind}
\end{subfigure}
\vspace{-.2cm}
\caption{\footnotesize{Performance on partial scans with overlap-aware (a-c) and overlap-blind (d) classifier training.}}
\label{fig:partial_acc}
\vspace{-.5cm}
\end{figure}

We evaluate the performance of our method with scan partiality.
We first focus on performance, where the training and testing of the classifier are overlap-aware, meaning that a different classifier is trained at each partial overlap. Fig.~\ref{fig:partial_acc_20db}-\ref{fig:partial_acc_4db} present modrec accuracy for 20, 10 and 4dB, respectively. Our method (LP+HOC) persistently outperforms existing counterparts across all SNR regimes. For a 100\% scan, our method performs on par with the literature for high SNR regimes (20 and 10dB) and outperforms the state in low SNR regimes (4dB). Table~\ref{tab:breakdown} (right) presents a breakdown of LP+HOC accuracy across modulations at SNR=20dB. Our method maintains high accuracy for low-order modulations even when a transmitter's bandwidth is scanned only at 50\%. The accuracy with high order modulations deteriorates as the overlap decreases. These results demonstrate that LP+HOC (i) leads to better modrec performance across all partial scans and (ii) is robust in noisy channel conditions.

Prior knowledge of the scan partiality in the classifier training phase poses a practical challenge to the real-world applicability of our method, as one needs to determine the fraction at which a transmitter is scanned before recognizing its modulation. Thus, we consider overlap-blind classification in which the classifier is trained on a mix of all possible partial scans as opposed to at every overlap individually. 
Fig.~\ref{fig:partial_acc_blind} shows our results for overlap-blind modrec at a challenging SNR of 10dB. Global features (HOC and OS) alone fail in classification even at a 100\% coverage. LP+HOC and LP, on the other hand, outperform their global counterparts across all partial overlap. Furthermore, in comparison with the the overlap-aware modrec performance at 10dB (Fig.~\ref{fig:partial_acc_10db}), our method only suffers a marginal performance deterioration when overlap-blind classifier training is employed. 
These results demonstrate our method's applicability in the wild without the need of prior knowledge of the partiality of a transmitter's scan.

\begin{table}[]
\center
\caption{\footnotesize{LP+HOC accuracy with scan bias (L) and partiality (R).}}\label{tab:breakdown}
\scriptsize
\begin{tabular}{c|c|c|c|c||c|c|c|c|c|c|}
\cline{2-9}
                             & \multicolumn{4}{c||}{Bias at 10dB, \#missing symbols} & \multicolumn{4}{c|}{Overlap at 20dB, \%} \\ \cline{2-9}
                             & 0         & 1         & 2         & 3        & 100    & 90    & 80    & 70    \\ \hline
\multicolumn{1}{|l|}{\texttt{\textbf{4-PSK}}}  &   1.00        &    1.00       &     0.87       &     1.00   &  1.00      &  1.00     &   1.00    &   0.99    \\ \hline
\multicolumn{1}{|l|}{\texttt{\textbf{8-PSK}}}  &   1.00        &  1.00         &    0.98        &    0.94    &  1.00      &  1.00     &   1.00    &   0.97   \\ \hline
\multicolumn{1}{|l|}{\texttt{\textbf{8-QAM}}}  &   1.00        &   0.97         &    0.83       &      0.84  &  1.00      &  1.00     &   1.00    &   1.00   \\ \hline
\multicolumn{1}{|l|}{\texttt{\textbf{16-QAM}}} &    0.77       &   0.66        &      0.65     &   0.61      &  1.00      &  0.81     &   0.64    &   0.63   \\ \hline
\multicolumn{1}{|l|}{\texttt{\textbf{64-QAM}}} &    0.74       &   0.74        &     0.71      &    0.70     &  1.00      &  0.86     &   0.64    &   0.56   \\ \hline
\end{tabular}
\end{table}

\vspace{-.4cm}
\subsection{Robustness to constellation rotation}\label{sec:rotation}

\begin{figure}[t]
\centering
\begin{subfigure}[t]{0.32\linewidth}
    \centering
    \includegraphics[width=\linewidth,trim={1cm 0 1cm 0},clip]{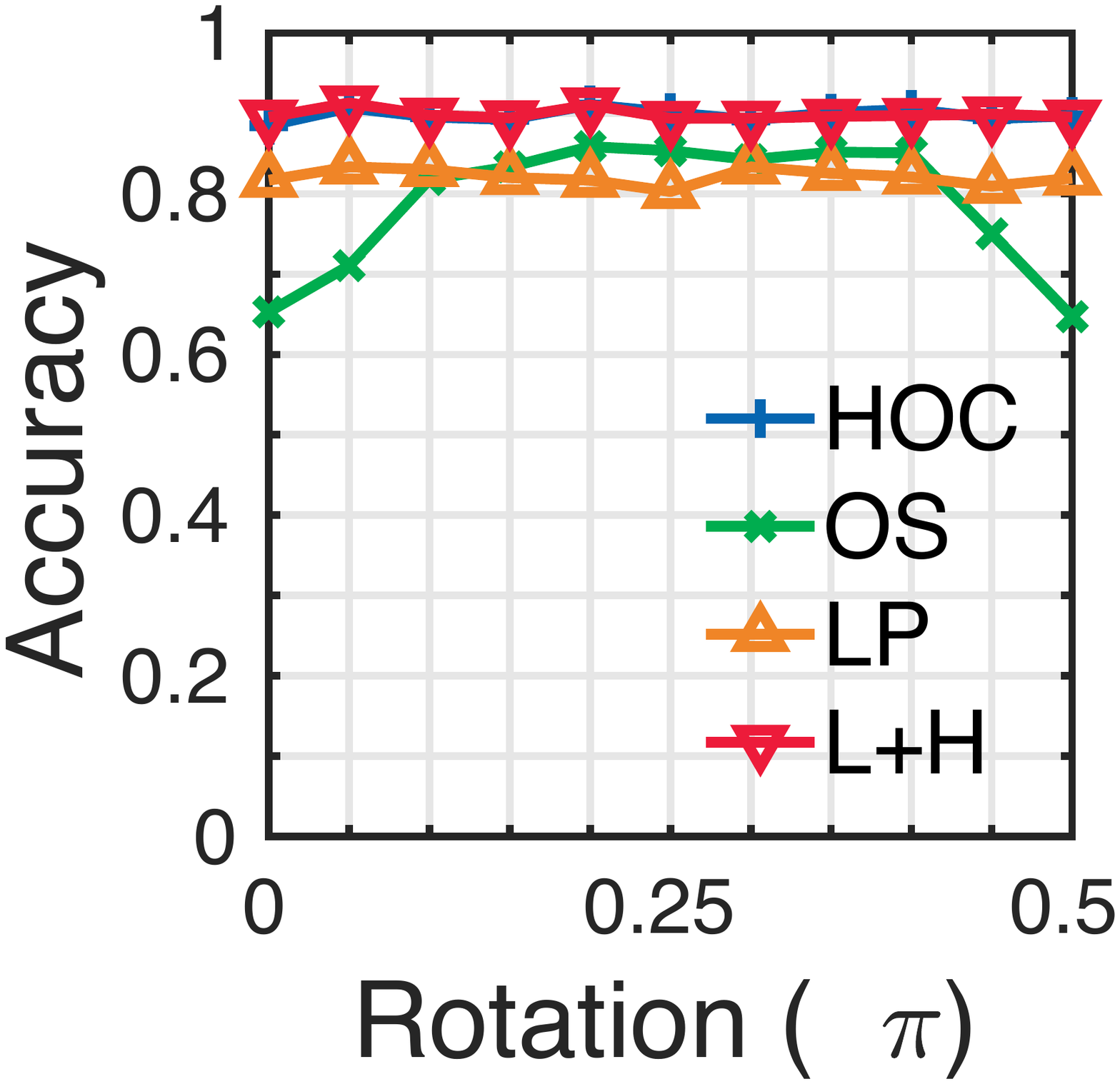}
    \caption{100\% overlap}\label{fig:phase100}
\end{subfigure}
\begin{subfigure}[t]{0.32\linewidth}
    \centering
    \includegraphics[width=\linewidth,trim={1cm 0 1cm 0},clip]{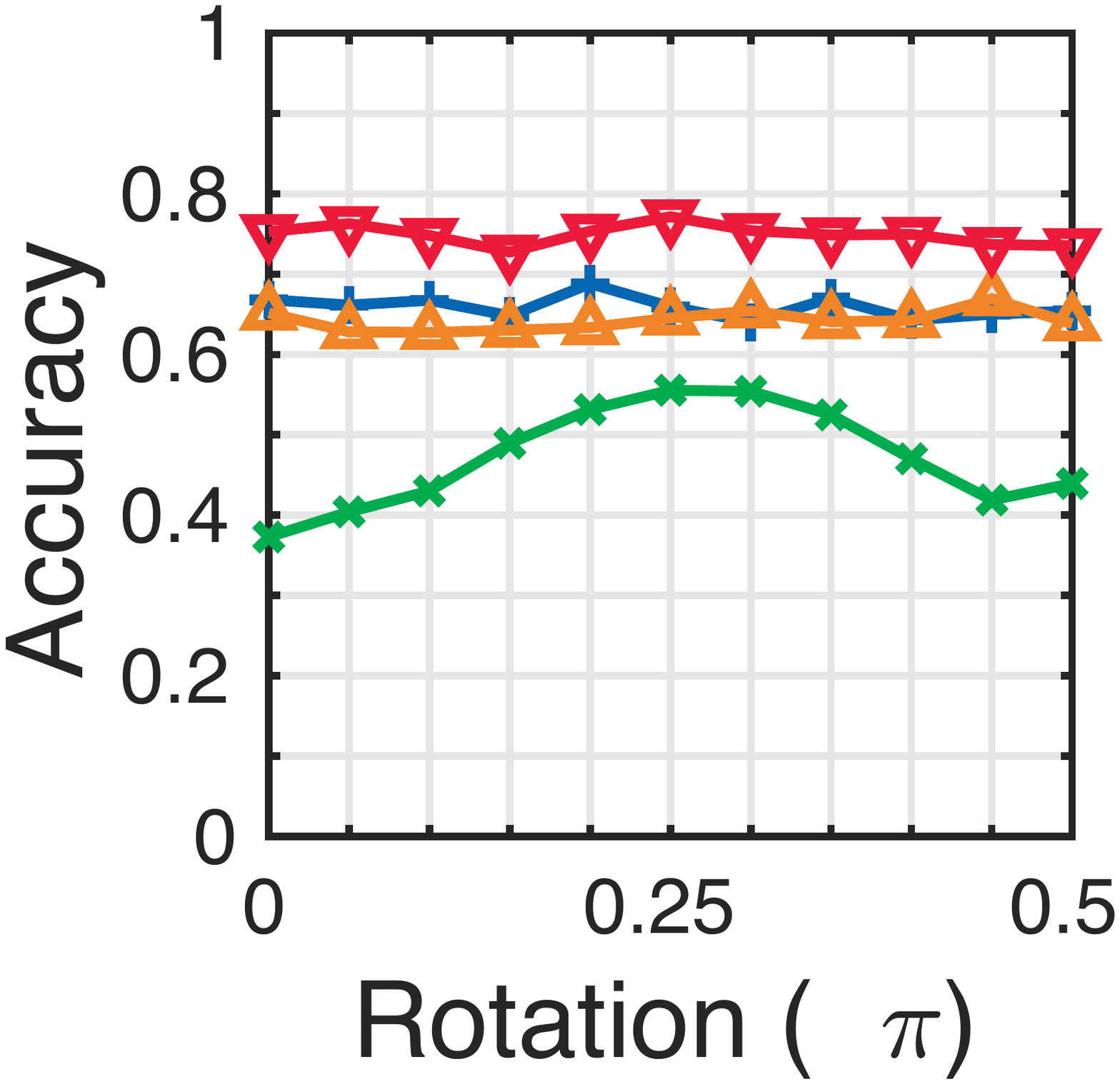}
    \caption{75\% overlap}\label{fig:phase75}
\end{subfigure}
\begin{subfigure}[t]{0.32\linewidth}
    \centering
    \includegraphics[width=\linewidth,trim={1cm 0 1cm 0},clip]{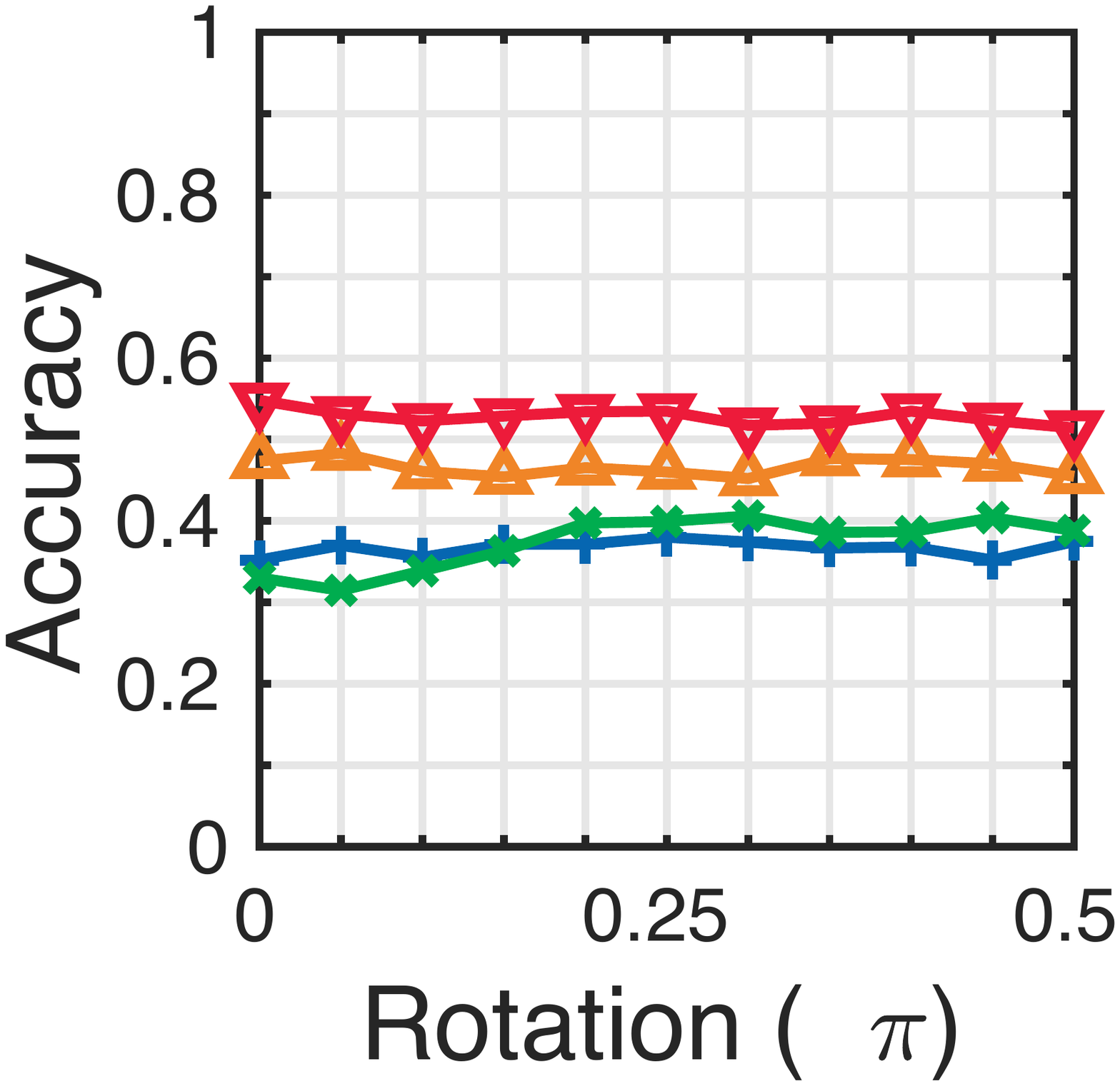}
    \caption{50\% overlap}\label{fig:phase50}
\end{subfigure}
\vspace{-.2cm}
\caption{\footnotesize{Phase-blind classification of non-biased scans at SNR 10dB.}}
\label{fig:rotation}
\end{figure}

We evaluate the performance of our approach with unknown constellation rotation. We use spectrum scans at a challenging SNR regime of 10dB. We vary the constellation rotation from 0 to $\pi/2$ while scanning a transmitter bandwidth at 100\%, 75\% and 50\%. We train our classifier on a mix of constellation rotations. Fig.~\ref{fig:rotation} presents our results. For 100\% scan HOC and LP+HOC perform equally well, whereas OS and LP have lower accuracy. As the transmitter overlap decreases, our method maintains maximal performance, whereas all other counterparts suffer dramatic deterioration in modrec accuracy due to their susceptibility to constellation rotation.

\vspace{-.4cm}
\subsection{Robustness to noise}\label{sec:noise}

All results so far were obtained with a SNR-aware classifier, meaning that a separate classifier was trained for each SNR level. This approach requires prior knowledge of the channel, which while feasible, adds steps and computational overhead to the modrec procedure. To address this issue, we explore SNR-blind modrec, where the classifier is trained on a mix of instances at different SNR levels. Specifically, we consider SNR levels from 0 to 20dB in increments of 2. At each SNR level we generate 1000 instances and train the classifier on the mix of these instances. We then test at each SNR level.

Fig.~\ref{fig:clf_accr_blind_snr} shows the classification accuracy across SNR. We compare our proposed method LP+HOC with HOC~\cite{swami2000hierarchical} and OS~\cite{lu2017modulation}. For the HOC features, we use all fourth order and sixth order cumulants. For the OS features, we use the amplitude and the phase order statistics. The same linear SVM classifier is used across all of HOC, OS and LP+HOC. The results show that our method is persistently able to achieve maximal performance across all SNR regimes. HOC alone suffers severe performance deterioration across all SNR levels, while OS performs on par with our method at high SNR and slightly worse in low SNR. 

\vspace{-.4cm}
\subsection{Effects of dictionary learning on modrec accuracy}\label{sec:dl}

So far, all experimental results were obtained with a single universally-trained dictionary as detailed in \S\ref{sec:data}. In this section we evaluate the feasibility of such a universal dictionary across various realistic scenarios. Of key interest is whether the training data (i.e. SNR level, scan overlap and bias) and parameter setting (i.e. dictionary and patch size) play a role in modrec performance. For this analysis, we generate a synthetic data set using our simulator (\S\ref{sec:data}) with 100\% transmitter overlap, no data bias and mixed constellation rotation. We consider ten SNR levels (0-20dB in increments of 2) and five modulations: \texttt{QPSK}, \texttt{8-PSK}, \texttt{8-QAM,} \texttt{16-QAM} and \texttt{64-QAM}. For each modulation and SNR level we generate $m=1000$ training instances of $n=512$ each and another $500$ testing instances of the same size. The SVM classifier was SNR-aware and trained on 100\% scans with no data bias.

\begin{figure}[t]
\centering
\begin{minipage}{.32\linewidth}
\centering
\includegraphics[width=.9\linewidth,trim={1cm 0 1cm 0},clip]{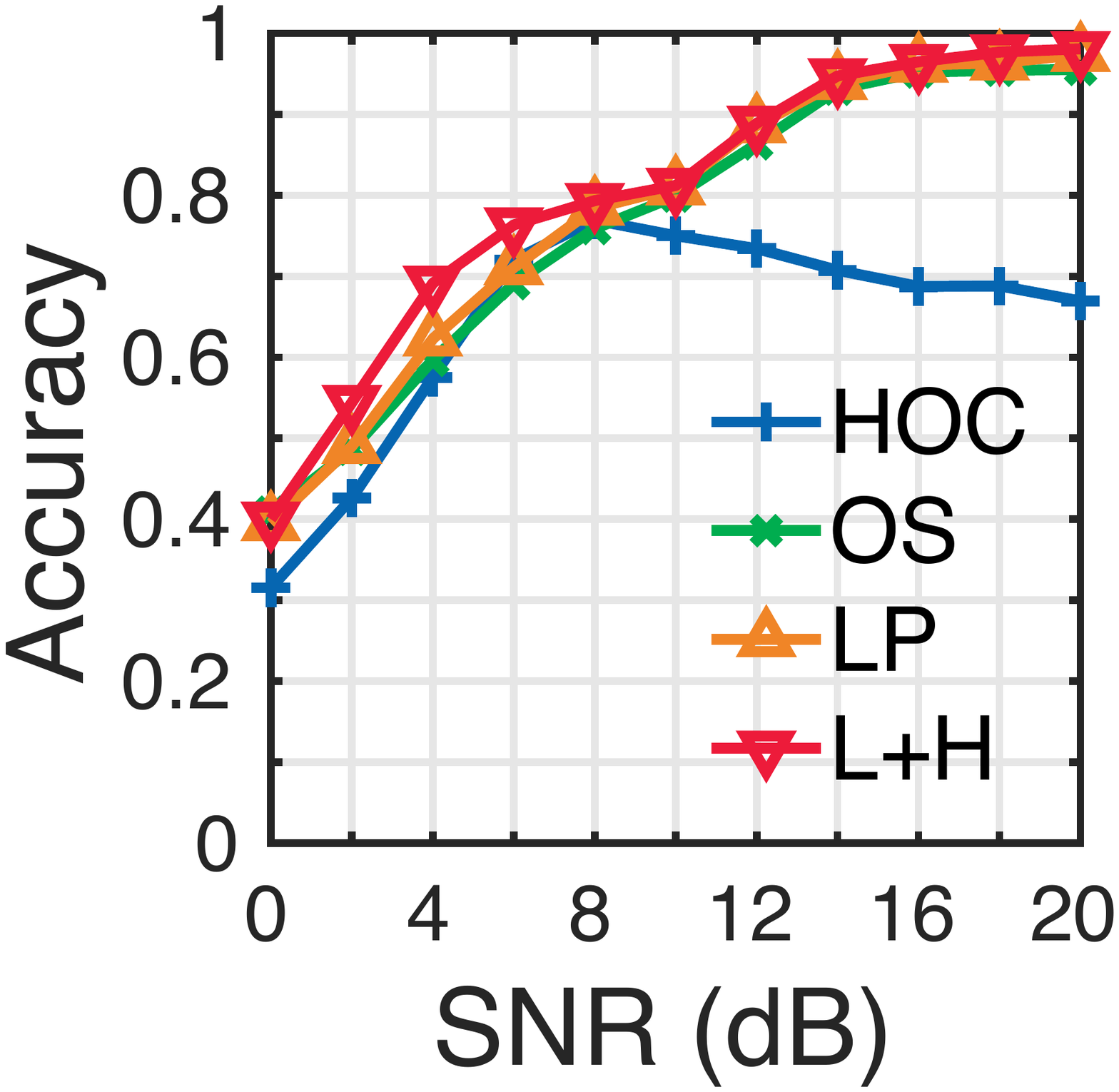}
\caption{\footnotesize{SNR-blind classification of 100\% scan overlap with no data bias.}}
\label{fig:clf_accr_blind_snr}
\end{minipage}
\begin{minipage}{.67\linewidth}
\centering
\begin{subfigure}[t]{0.49\linewidth}
    \centering
    \includegraphics[width=\linewidth]{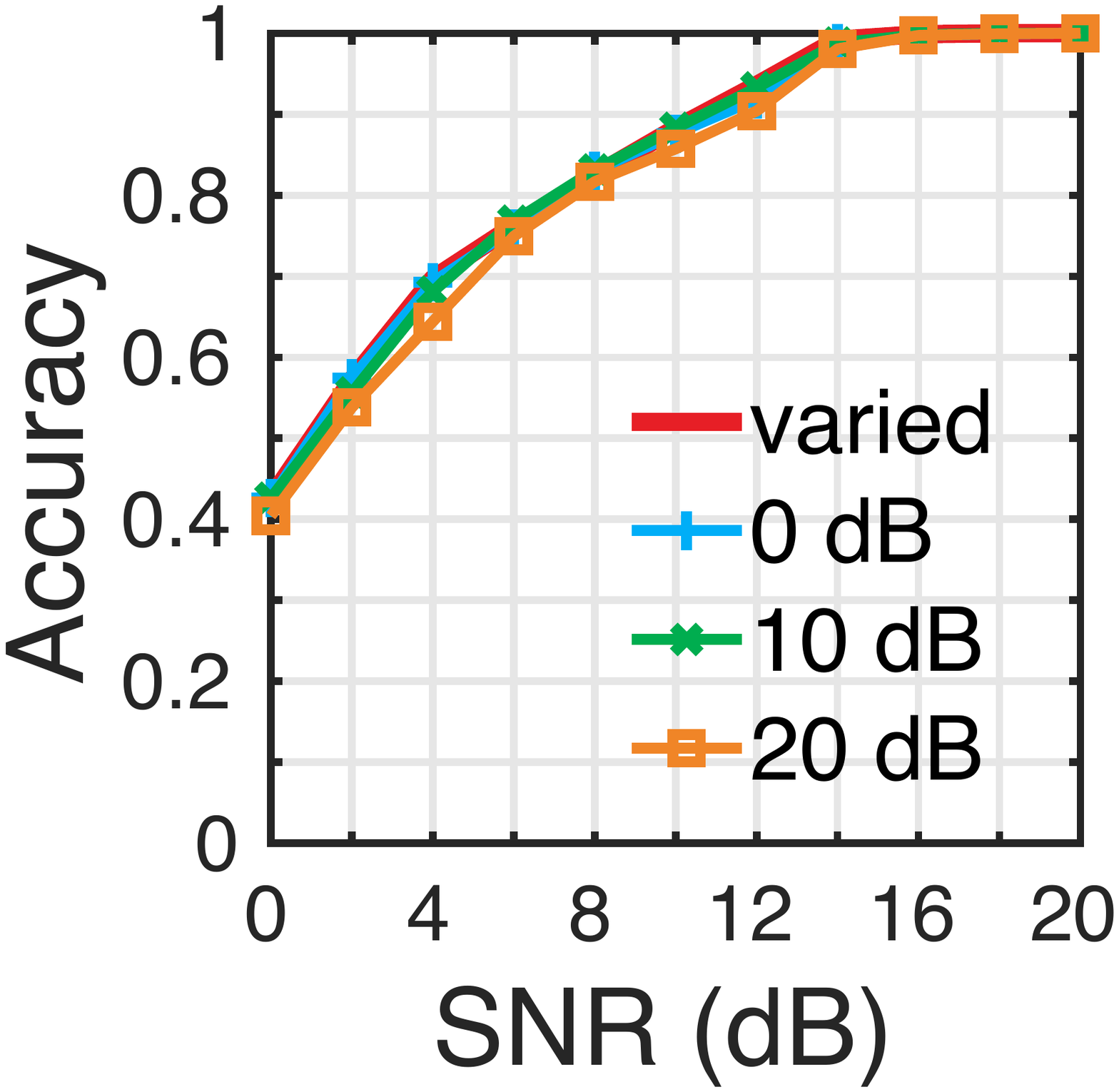}\label{fig:accr_dict_vs_snr}
\end{subfigure}
\begin{subfigure}[t]{0.49\linewidth}
    \centering
    \includegraphics[width=\linewidth]{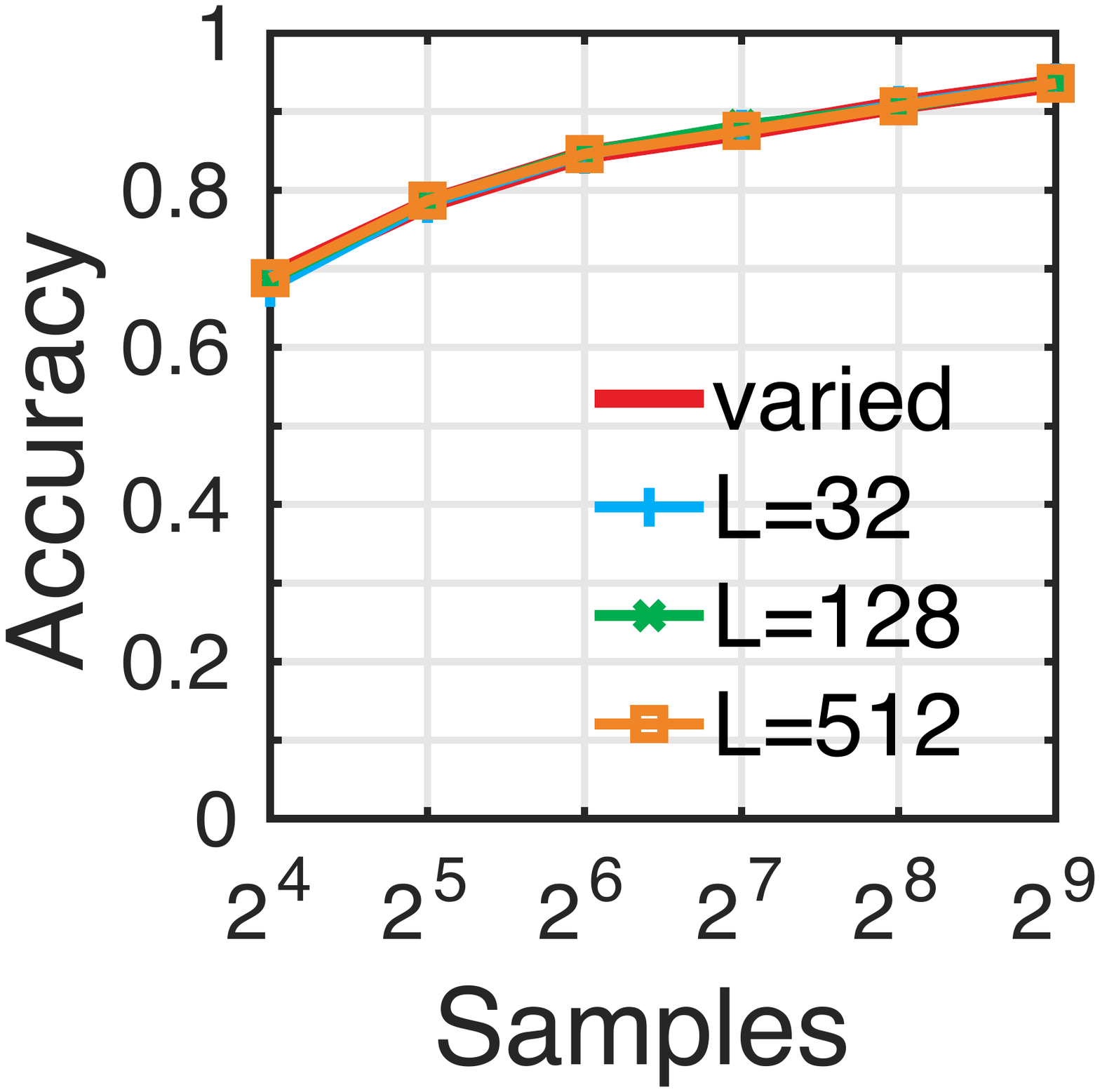}\label{fig:accr_dict_vs_nlen}
\end{subfigure}
\caption{\footnotesize{Dictionary learning performance over SNR (left) and instance size (right) for four training mechanisms. DL
    does not require explicit training for each SNR or instance size.}}\label{fig:dlperf}
\end{minipage}
\end{figure}

\noindent \textbf{1) Should dictionary learning be SNR-, overlap-, and bias-aware?} We adopt four training approaches in the dictionary learning phase: (1) \textit{Varied}: we train and test the dictionary for each SNR level, (2) \textit{0dB}: we train the dictionary at SNR=0dB and test at each level, (3) \textit{10dB}: we train the dictionary at SNR=10dB and test at each level and (4) \textit{20dB}: we train the dictionary at 20dB and test at each level. Fig.~\ref{fig:dlperf} (left) presents modrec accuracy for the four training strategies across a range of SNR values from 0 to 20dB. The classification performance remains the same across training approaches, which indicates that we can only train the dictionary once, at any SNR level and the learned patterns will be applicable across any SNR level. Similar conclusions can be drawn from our results with biased and partially-overlapping scans (results omitted in interest of space). This is particularly important to the real world applicability of our approach, as it demonstrates its robustness to various real-world conditions.

\noindent \textbf{2) Does the instance length affect the dictionary learning performance?} The instance length is defined in terms of the number of \iq samples that appear in a measured sequence. We adopt four dictionary training approaches: (1) \textit{Varied}: we train and test the dictionary for each instance length, (2) \textit{L=32}: we train the dictionary at instance length of 32 and test at all instance lengths, (3) \textit{L=128}: we train the dictionary at instance length of 128 and test at all instance lengths and (4) \textit{L=512}: we train the dictionary at instance length 512 and test on all instance lengths. Fig.~\ref{fig:dlperf} (right) presents modrec accuracy over increasing instance length for the four training schemes. We see that the classification performance remains the same across the four training approaches, indicating that we do not need to retrain the dictionary as our instance length changes.

\noindent \textbf{3) Do dictionary learning parameters affect the modrec accuracy?} Two key parameters of the dictionary learning step are the shingle size $l$ and the number of components $K$ for the GMM instantiation. 
We explore the performance with various $(l,K)$ combinations ($l={2, 3, 5}$, $K=20,50,100$) with SNR varying from 0 to 10dB in increments of 2. In interest of space we omit a figure and summarize our results as follows. We observe maximal performance across all SNRs for shingle size of 2 or 3, which deteriorates at $l=5$. Similarly, a GMM instantiation with 20 or 50 components leads to good accuracy, however, the accuracy deteriorates with $K=100$. Thus, we choose to use a shingle size of $3$ and a dictionary size of $50$.

\vspace{-.3cm}
\subsection{Discussion}
Our evaluation shows the merit of combining local and global features for robust modrec in the wild. Our hierarchical modrec approach LP+HOC outperforms methods based on global features by a large margin across all the explored realistic scenarios. A counterpart based on LP-only closely follows the performance of LP+HOC, however, it is not as robust in the face of arbitrary constellation rotation.

\vspace{-.2cm}
\section{\ourmeth on partial scans in the wild}\label{sec:eval_real}

\begin{figure}
    \centering
    \begin{subfigure}{.15\textwidth}
   \includegraphics[width=\linewidth]{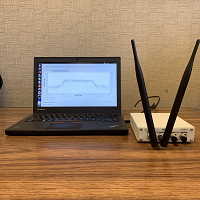}
    \caption{Transmitter}
    \label{fig:tx_setup}
     \end{subfigure}
     \begin{subfigure}{.15\textwidth}
   \includegraphics[width=\linewidth]{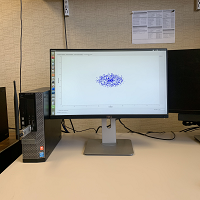}
    \caption{RTL}
    \label{fig:rtl_setup}
     \end{subfigure}
      \begin{subfigure}{.15\textwidth}
   \includegraphics[width=\linewidth]{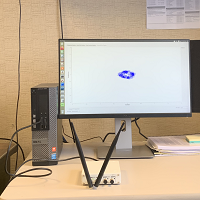}
    \caption{USRP}
    \label{fig:rtl_setup}
     \end{subfigure}
     \caption{\footnotesize Sensor Platforms Testbed Setup}
     \label{fig:testbed}
     \vspace{-.5cm}
\end{figure}

\begin{figure*}[t]
\begin{minipage}{\linewidth}
\centering
\begin{subfigure}{.195\textwidth}
\includegraphics[width=\linewidth]{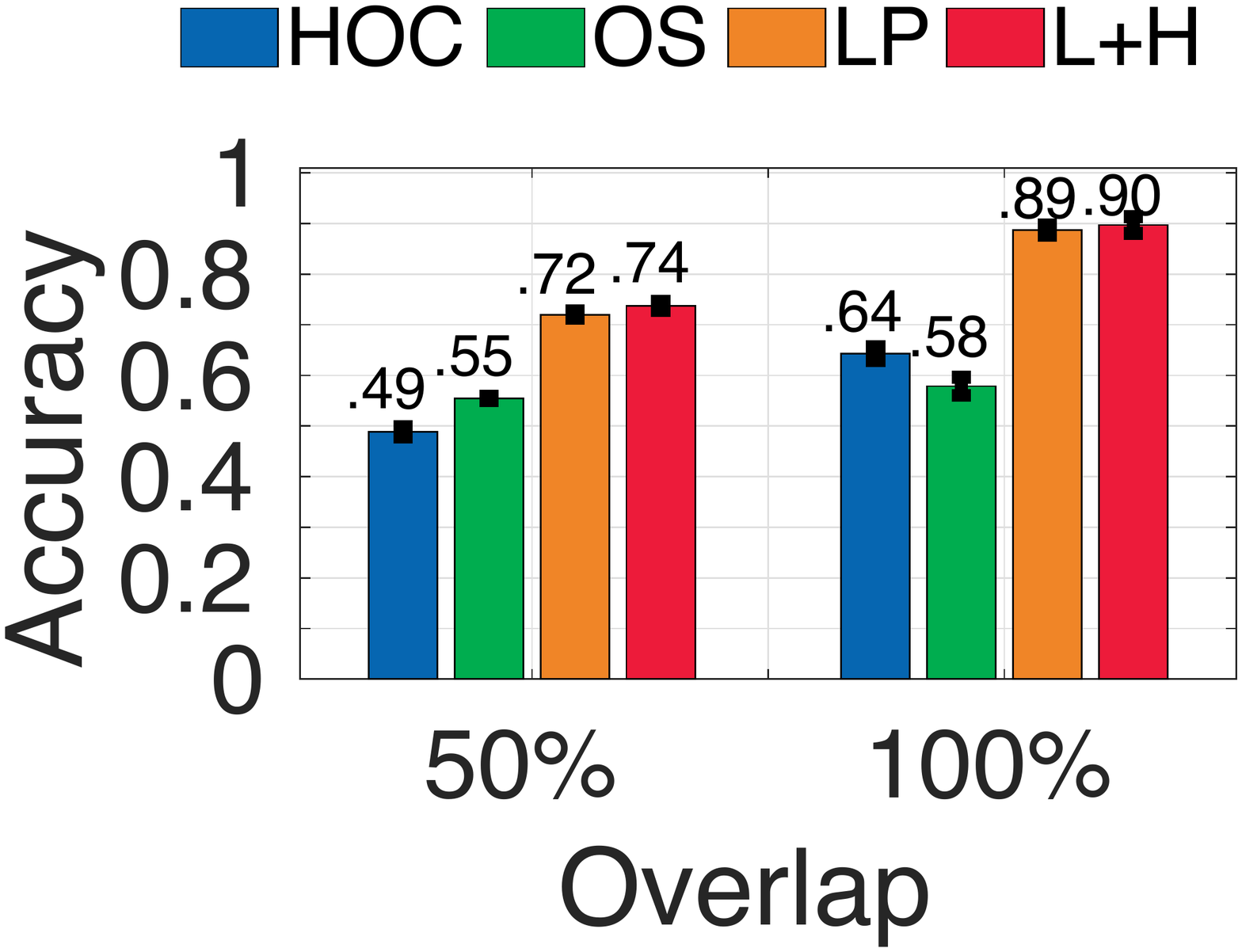}
\caption{RTL}\label{fig:rtl-rtl}
\end{subfigure}
\begin{subfigure}{.195\textwidth}
\includegraphics[width=\linewidth]{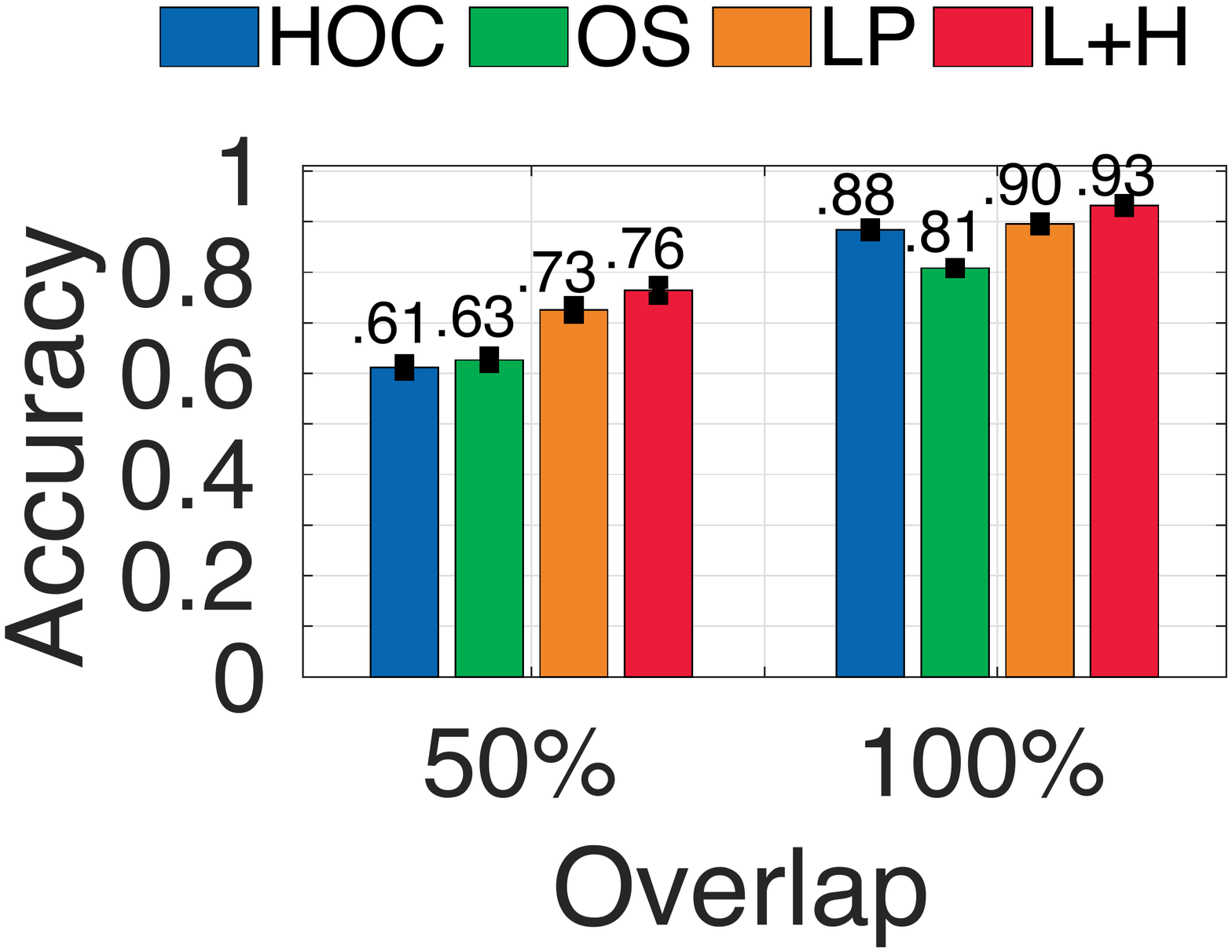}
\caption{USRP}\label{fig:usrp-usrp}
\end{subfigure}
\begin{subfigure}{.195\textwidth}
\includegraphics[width=\linewidth]{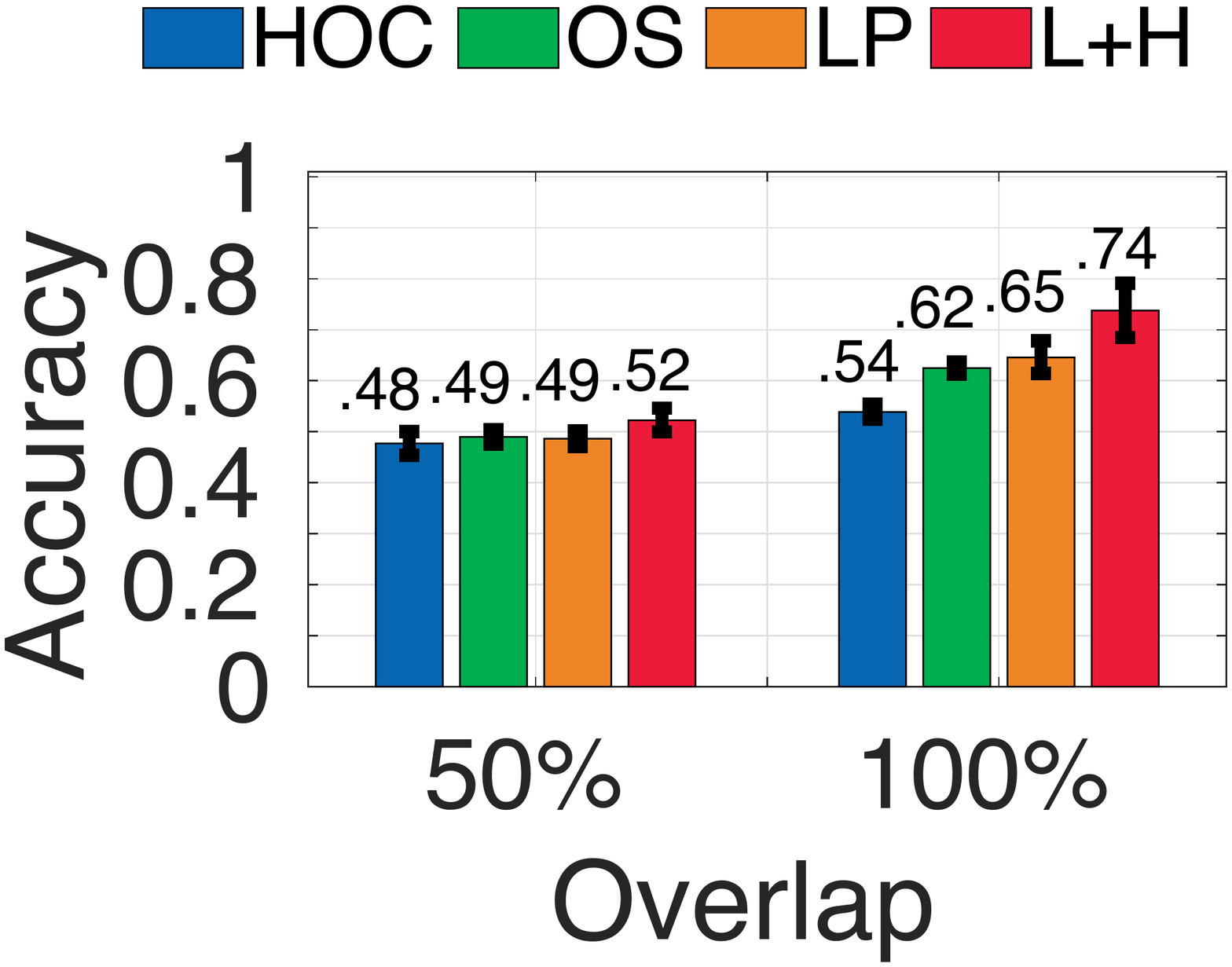}
\caption{RTL $\rightarrow$ USRP}\label{fig:rtl-usrp}
\end{subfigure}
\begin{subfigure}{.195\textwidth}
\includegraphics[width=\linewidth]{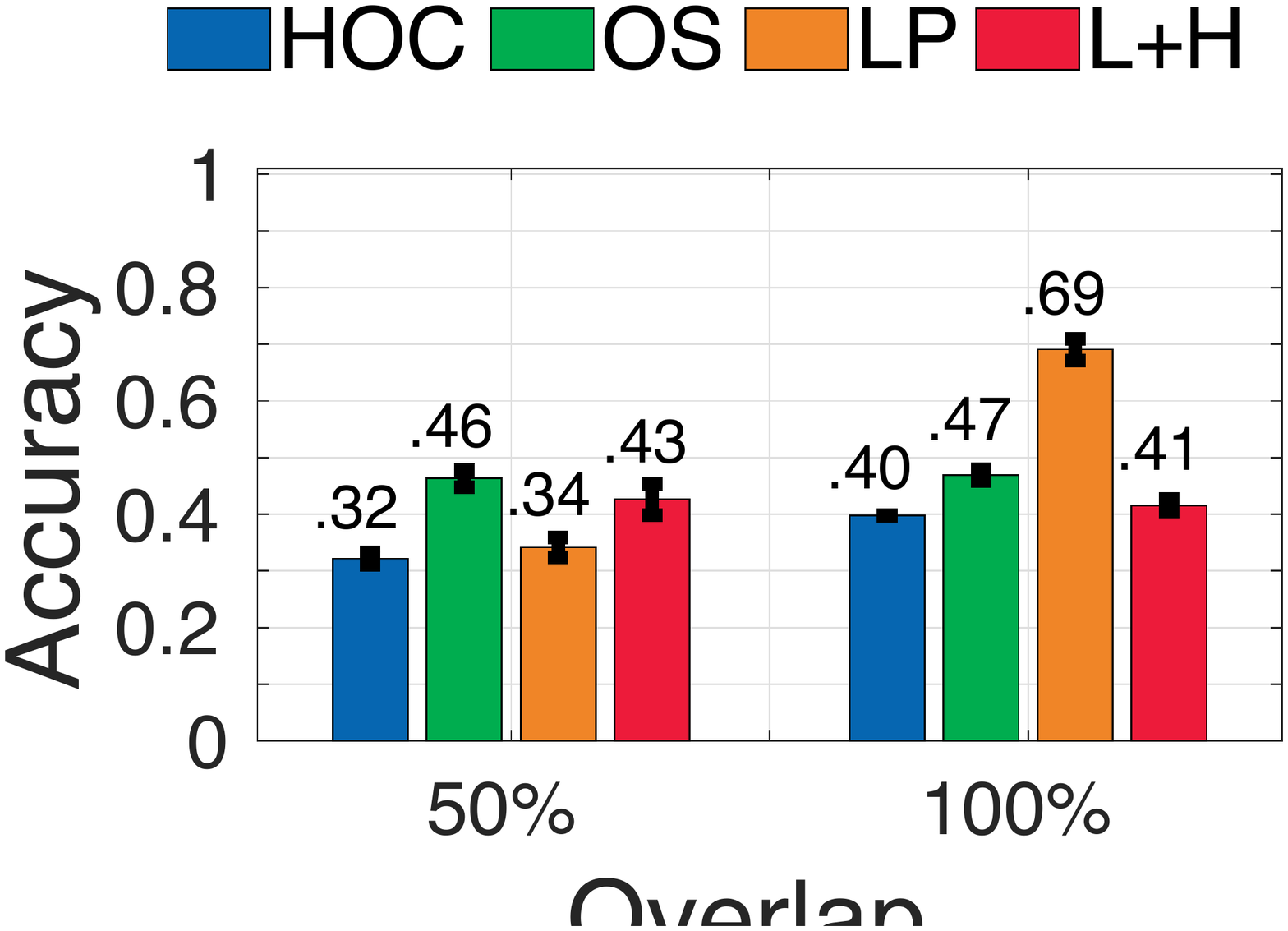}
\caption{USRP $\rightarrow$ RTL}\label{fig:usrp-rtl}
\end{subfigure}
\begin{subfigure}{.195\textwidth}
\includegraphics[width=\linewidth]{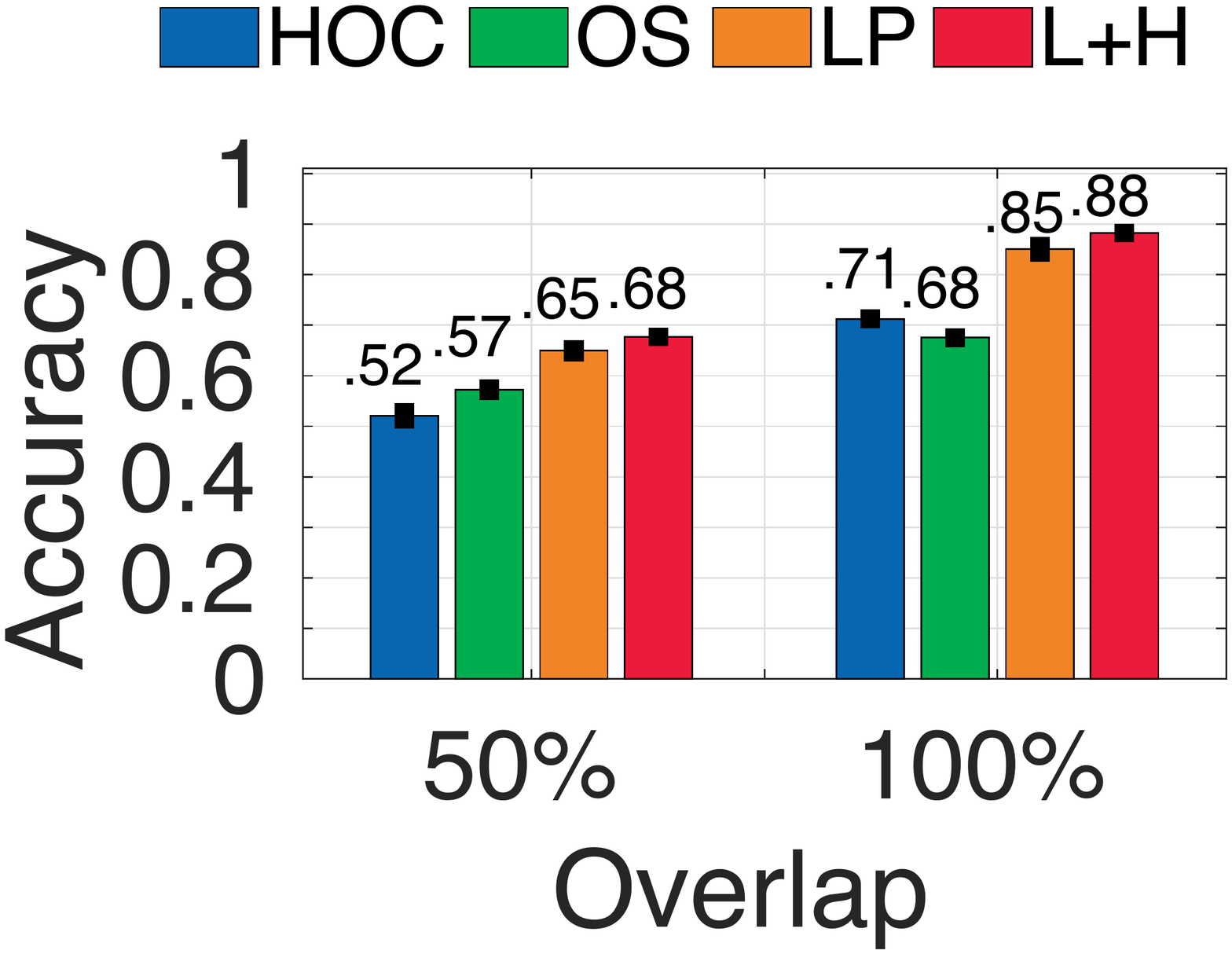}
\caption{Mixed}\label{fig:mixed-mixed}
\end{subfigure}
\vspace{-.2cm}
\caption{\footnotesize{Modrec from partial scans in a software-defined radio testbed. }}\label{fig:ota_results_2019}
\end{minipage}
\vspace{-.5cm}
\end{figure*}

In this section we evaluate the performance of our method on real over-the-air transmissions from a software-defined radio testbed. Beyond classification accuracy we are also interested in \ourmeth's performance across sensor platforms. To this end, we employ two types of sensors: one using a \$1,500 USRP B210 and an other using a \$20 RTL-SRD as a radio frontend. In what follows we first describe our experimental setup. We then discuss performance across sensor platforms. Finally, we evaluate and discuss the dependency of modulation classification on platform-aware training.  

\vspace{-.2cm}
\subsection{Experimental setup.} We collect data from controlled transmissions in a testbed comprised of two sensors and one transmitter as pictured in Fig.~\ref{fig:testbed}. The transmitter was comprised of a USRP B210 attached to a Laptop with Intel i7-5600U CPU and 8GB of RAM. The first sensor used an RTL-SDR connected to a PC with an Intel i7-4770 CPU and 16GB RAM, whereas the second sensor was comprised of a USRP B210 attached to a PC with Intel i7-4770 CPU and 16GB RAM. For each experiment, the transmitter and receiver were located in line of sight at a distance of roughly 5 feet. We set the transmitter gain at 65dBm, the receiver gain for the RTL at 40dBm and the USRP at 50dBm. The transmitter uses GNURadio \cite{gnuradio} to generate a signal modulated with \texttt{BPSK}, \texttt{QPSK}, \texttt{8PSK}, \texttt{QAM16}, and \texttt{QAM64}. We evaluate the performance of our modrec by collecting scans at 50\% and 100\% transmitter overlap for each sensor. 
For this experiment, we set the center frequency of both the transmitter and the sensors to 1.2 GHz.
The instantaneous bandwidth of the transmitter was set to 2MHz, and its modulation was varied across the five types discussed above. The sensors' bandwidth was set to 2MHz for the 100\% coverage setting and 1MHz for the 50\% coverage setting.

\vspace{-.4cm}
\subsection{\ourmeth performance with platform-aware training}\label{sec:patform-aware}

We begin by evaluating \ourmeth's performance with platform-aware training, that is, both the training and the modulation classification were performed with traces collected from the same sensor platform. For this experiment we generated $1,000$ instanced of each of the five modulations and on each of the two sensor platforms (i.e. $5,000$ instances altogether for each platform). Each instance contained 128 samples. We report results as an average over 5 runs. For each run, we used 80\% of the instances for training and the remaining 20\% for testing. The training-testing splits were performed arbitrarily for each of the five runs. Fig.~\ref{fig:rtl-rtl} and \ref{fig:usrp-usrp} present the results from the RTL-SDR and the URSP sensors, respectively. The x-axis captures the overlap percentage, whereas the y-axis presents the classification accuracy. Blue and green present existing counterparts using high order cumulants (HOC) and order statistics (OS). Orange and red present two versions of \ourmeth: one that uses only local patterns for classification (labeled as LP) and another that uses a combination of local patterns and high order statistics (labeled as L+H). 

There are two key observations to be made. First, at 50\% overlap, the two versions of \ourmeth achieve 72/74\% accuracy with the RTL sensor and 73/76\% accuracy for the USRP sensor. This constitutes an average improvement of 43\% over existing features for the RTL sensor and 24\% improvement for the USRP sensor. \textit{In summary, \ourmeth's performance with partial scans is high and consistent compared to existing counterparts for both RTL- and USRP-based sensors.} In addition, we observe significant advantages of \ourmeth at full scan overlap with low-cost RTL-SDR sensors compared to existing counterparts. Focusing on the results for 100\% overlap in Fig.~\ref{fig:rtl-rtl}, we note that both versions of \ourmeth outperform HOC by 41\% and OS by 55\%. \textit{This result demonstrates the potential of careful feature design to support extremely low-cost transmitter fingerprinting, which was considered impossible to date.}

\vspace{-.4cm}
\subsection{\ourmeth performance with cross-platform and mixed-platform training}

Finally, we evaluate \ourmeth's performance with cross-platform and mixed-platform training. We seek to answer whether traces collected on one platform can be used to support modulation classification on a different platform. This is important for practical modrec, with heterogeneous sensor capabilities, since unified training and testing might not be always possible. For this experiment we use the same $10,000$ instances generated for the platform-aware evaluation (\S\ref{sec:patform-aware}). We present cross-platform training results for two settings: (i) training on RTL-SDR and testing on USRP (Fig.~\ref{fig:rtl-usrp}) and (ii) training on USRP and testing on RTL-SDR (Fig.~\ref{fig:usrp-rtl}). In addition we also evaluate \ourmeth while training and testing on a mix of traces from RTL-SDR and USRP. In each of the cross-platform settings, we use 80\% of the training traces and 20\% of the testing traces (e.g. for the RTL$\rightarrow$USRP case, we use $4,000$ RTL-SDR instances for training and $1,000$ USRP instances for testing). In the mixed-platform setting we use 80\% of (i.e. $8,000$ instances) for training and 20\% (i.e. $2,000$ instances) for testing. We present average over five runs. The training and testing subsets were randomly drawn for each run.

Fig.~\ref{fig:rtl-usrp}, \ref{fig:usrp-rtl} and \ref{fig:mixed-mixed} present our results. As in the platform-aware experimentation, blue and green present HOC and OS-based counterparts from the literature, whereas orange and red present two versions of \ourmeth. Several observations can be made. First, for cross-platform training at 50\% scan overlap (Fig.~\ref{fig:rtl-usrp} and \ref{fig:usrp-rtl}), both RTL$\rightarrow$USRP and USRP$\rightarrow$RTL perform worse than their platform-aware counterparts. Furthermore, all methods, both existing and our proposed, achieve similar performance. At 100\% overlap variants of \ourmeth outperform HOC- and OS-based counterparts (i.e. \ourmeth with LP achieves 74\% accuracy in the RTL$\rightarrow$USRP case, while \ourmeth with HOC+LP achieves 70\% accuracy in the USRP$\rightarrow$RTL case). This result indicates the need for further investigation in counterpart selection with cross-platform training. Finally, our results from the mixed pool training and testing (Fig.~\ref{fig:mixed-mixed}) achieve accuracy commensurate with that of the platform-aware training. \textit{In summary, our results indicate that in cases where platform-aware training is not possible, a mixed-pool training may facilitate high performance, however, cross-platform training would lead to poor modrec performance. }

\vspace{-.3cm}
\section{Conclusion} \label{sec:conclusion}

We designed a novel modulation classification framework dubbed \ourmeth, which is robust to imperfect spectrum scan data due to encoding local sequential patterns within \iq samples. We employed a Fisher Kernel representation which flexibly handles non-linearity in the underlying data and enables high-quality modulation recognition even with simple linear classification models such as linear soft-margin Support Vector Machines. We demonstrated our framework's applicability on real-world, partial, intermittent, biased and noisy scans. Our method consistently outperformed state-of-the-art approaches, and in addition our local features were demonstrated to encode complementary information to global alternatives. Thus, our framework can be effectively combined with existing features to further boost its individual recognition accuracy.

Our work addresses a critical disconnect between modrec requirements and spectrum sensing capabilities. Particularly, it addresses critical challenges posed by emerging spectrum-sharing technologies which will employ heterogeneous dedicated or crowdsourced sensors, scanning a wide frequency band sequentially, and thus producing intermittent, partial and noisy scans. The superior performance of our methodology on over-the-air partial scans from $\$1,500$ USRP and $\$20$ RTL-SDR indicates its potential for improved modrec in the wild. In addition, our RTL-based evaluation demonstrates  the  potential  of  careful  feature  design  to support extremely low-cost transmitter fingerprinting, which was considered impossible to date. Our proposed framework and its inter-operability with previous approaches constitutes a solid foundation for future work on spectrum analytics with practical importance to future spectrum sharing technology, enforcement and security.

\section{Acknowledgements}
This work was supported through NSF CISE Research Initiation Initiative (CRII) grant CNS-1657476 and NSF CAREER grant CNS-1845858.


%




\ifCLASSOPTIONcaptionsoff
  \newpage
\fi




\bibliographystyle{abbrv}
\bibliography{main}
%



%

\begin{IEEEbiography}[{\includegraphics[width=1in,clip,keepaspectratio]{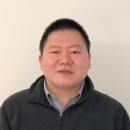}}]{Wei Xiong}
is working toward the PhD degree in the Department of Computer Science, University at Albany. His research focus is on modulation recognition. His work spans a wide range of topics including novel machine learning methods in modulation recognition applications, small-scale cellular networks systems, and network connectivity in resource-limited environments.
\end{IEEEbiography}

\vspace{-1cm}
\begin{IEEEbiography}[{\includegraphics[width=1in,height=1.25in,clip,keepaspectratio]{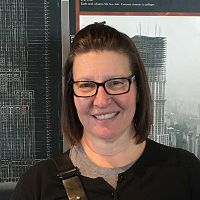}}]{Karyn Doke}
is working toward the PhD degree in the Department of Computer Science, University at Albany. Her research interests are in automated spectrum management and the development of machine learning algorithms to characterize spectrum. She also works with wireless networks in rural communities to improve emergency preparedness services.
\end{IEEEbiography}

\vspace{-1cm}
\begin{IEEEbiography}[{\includegraphics[width=1in,height=1.25in,clip,keepaspectratio]{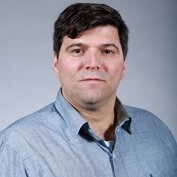}}]{Dr. Petko Bogdanov} is an Assistant Professor at the computer science department of University at Albany -- SUNY. His research interests include data mining and management and applications to bioinformatics, neuroscience, data-driven nanomaterial design and wireless networks. Previously, he was a postdoctoral fellow at the department of computer science at the University of California, Santa Barbara. He received his PhD and MS in Computer Science from the University of California, Santa Barbara in 2012 and his BE in Computer Engineering from Technical University---Sofia in 2005. Dr. Bogdanov is a member of the IEEE and the ACM and his research has been supported by grants from NSF, DARPA and ONR. He currently serves as an Associate Editor for IEEE Transactions on Knowledge and Data Engineering (TKDE).
\end{IEEEbiography}

\vspace{-1cm}
\begin{IEEEbiography}[{\includegraphics[width=25mm,height=25mm,clip,keepaspectratio]{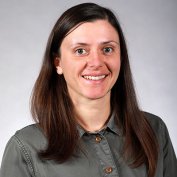}}]%
{Dr. Mariya Zheleva}
is an Assistant Professor in the Department of Computer Science at University at Albany -- SUNY. Her research is at the intersection of wireless networks and Information and Communication Technology for Development. She has done work on small local cellular networks, data-driven Dynamic Spectrum Access, spectrum management and sensing and network performance and characterization. She is the founder and director of the UbiNet Lab at University at Albany. She holds a PhD and MS in Computer Science from the University of California, Santa Barbara and a M.Eng. and B.Eng. in Telecommunications from the Technical University -- Sofia. Her research is funded by the National Science Foundation and Microsoft. 
\end{IEEEbiography}






\end{document}